\documentclass[aps,prb,notitlepage,superscriptaddress,reprint]{revtex4-1}
\pdfoutput=1
\usepackage{amsfonts}
\usepackage{amsmath}
\usepackage{amssymb}
\usepackage{graphicx}
\usepackage{bm}
\usepackage{color,soul}
\usepackage{float}
\usepackage{indentfirst}

\begin{document}

\title{Tunable spin-orbit coupling in two-dimensional InSe}
\author{A. Ceferino}
\email{adrian.ceferino@postgrad.manchester.ac.uk}
\address{Department of Physics and Astronomy, University of Manchester, Oxford Road, Manchester, M13 9PL, United Kingdom}
\address{National Graphene Institute, Booth Street East, Manchester, M13 9PL, United Kingdom}
\author{S.J. Magorrian}
\address{National Graphene Institute, Booth Street East, Manchester, M13 9PL, United Kingdom}
\author{V. Z\'{o}lyomi}
\address{STFC Hartree Centre, Daresbury Laboratory,
Daresbury, Warrington, WA4 4AD, United Kingdom}
\author{D.A. Bandurin}
\address{Department of Physics, Massachusetts Institute of Technology,
Cambridge, Massachusetts 02139, USA}
\address{Department of Physics and Astronomy, University of Manchester, Oxford Road, Manchester, M13 9PL, United Kingdom}
\author{A.K. Geim}
\address{Department of Physics and Astronomy, University of Manchester, Oxford Road, Manchester, M13 9PL, United Kingdom}
\address{National Graphene Institute, Booth Street East, Manchester, M13 9PL, United Kingdom}
\author{\\ A. Patan\`{e}}
\address{School of Physics and Astronomy, University of Nottingham, Nottingham,  NG7 2RD, UK}
\author{Z.D. Kovalyuk}
\address{Institute for Problems of Materials Science, The National Academy of Sciences of Ukraine, Chernivtsi, 58001, Ukraine}
\author{Z.R. Kudrynskyi}
\address{School of Physics and Astronomy, University of Nottingham, Nottingham,  NG7 2RD, UK}
\author{I.V. Grigorieva}
\address{Department of Physics and Astronomy, University of Manchester, Oxford Road, Manchester, M13 9PL, United Kingdom}
\address{National Graphene Institute, Booth Street East, Manchester, M13 9PL, United Kingdom}
\author{V.I. Fal'ko}
\address{Department of Physics and Astronomy, University of Manchester, Oxford Road, Manchester, M13 9PL, United Kingdom}
\address{National Graphene Institute, Booth Street East, Manchester, M13 9PL, United Kingdom}
\address{Henry Royce Institute for Advanced Materials, Manchester, M13 9PL, United Kingdom}

\begin{abstract}
We demonstrate that spin-orbit coupling (SOC) strength for electrons near the conduction band edge in few-layer $\gamma$-InSe films can be tuned over a wide range. This tunability is the result of a competition between film-thickness-dependent intrinsic and electric-field-induced SOC, potentially, allowing for electrically switchable spintronic devices. Using a hybrid $\mathbf{k\cdot p}$ tight-binding model, fully parameterized with the help of density functional theory computations, we quantify SOC strength for various geometries of InSe-based field-effect transistors. The theoretically computed SOC strengths are compared with the results of weak antilocalization measurements on dual-gated multilayer InSe films, interpreted in terms of Dyakonov-Perel spin relaxation due to SOC, showing a good agreement between theory and experiment.
\end{abstract}

\maketitle
\section{Introduction}\label{section:Introduction}

Indium selenide (InSe) is a layered semiconductor with already demonstrated high mobility and versatile optical properties\cite{hammerindirect,tunablegap,PLMorpurgo,TerryInfraredtovioletTO,PhotodetectorLee,twistindir,Resonanttunel,tunelJohanna}. Atomically thin InSe films (exfoliated from bulk crystals \cite{exfoliated} or produced by chemical vapour deposition\cite{CVD}) have already been used to fabricate field-effect transistors (FET devices). Moreover, the persistence of high mobility\cite{bandurin2017high,mobilityInSe,highmobilityInSe2} of electrons in \emph{n}-type doped  $\gamma$-InSe to only few atomic layers \cite{PLMorpurgo,TerryInfraredtovioletTO,exfoliated} in thickness makes it feasible to implement InSe in spintronic devices\cite{monorashba}. In contrast to the conventional InAs\cite{InAs} or HgTe\cite{HgTe} quantum wells, in $\gamma$-InSe it is possible to completely suppress the conduction band SOC using applied displacement field, allowing for exceptionally tunable spintronic devices. One of the methods to control the electron spin in semiconductors is to manipulate its spin-orbit coupling (SOC)\cite{MorpurgoWL,BartWSe,WAL_Te,Review,jeanie,1DSoc,bilayerSOC}, and, in this paper, we study the dependence of SOC for two dimensional (2D) electrons near the conduction band edge of InSe films on the number of layers and on the gate-controlled electrostatic doping in the films implemented in the FET geometry\cite{reviewpremasiri,actaslovaka,AndradaSilva,Lossinterusbband,kpLassnig,FabianPhosphorene,VanderWaalsmanipulation,QWsubbandsinversion}.
\begin{figure}[t!]
    \includegraphics[width=\columnwidth,height=4.2in]{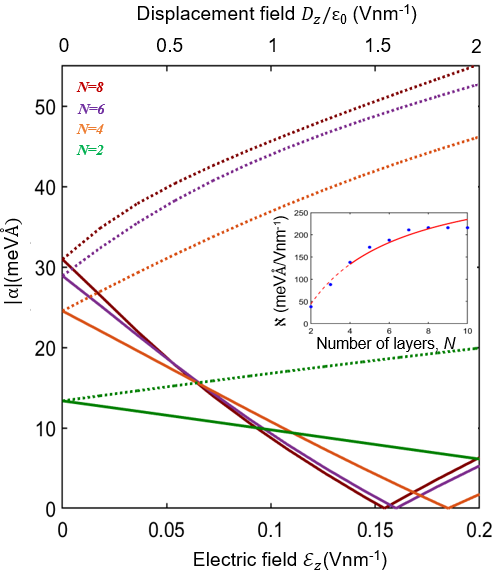}
	\caption{\footnotesize SOC strength dependence on displacement field and number of layers \textit{N} at $n_{e}=0$ in a dual-gated FET geometry. InSe dielectric constant\cite{dielectric} used here is $\epsilon_{z}=9.9$. The inset shows the layer-number-dependence of parameter $\aleph$ used to take into account the influence of an electric field $\mathcal{E}_{z}$ in Eq. (\ref{eqn:formSOCexp}). The solid and dotted lines indicate respectively when the applied displacement field suppresses or enhances the Dresselhaus SOC.}
	\label{fig:RashbaEfield}
\end{figure}\newline
\indent Below, we use the earlier developed hybrid $\bold{k\cdot p}$ tight-binding (HkpTB) model for InSe\cite{magorrian2016electronic,SamSpringer}, taking into account the s and $p_{z}$ orbital composition of the lowest conduction subband and self-consistent analysis of the electrostatic potential on each layer\cite{subbands2018}, and show that the dominant term in the SOC in $\gamma$-stacked InSe multilayer thin film (any number of layers) has the generic form,
\begin{align}
    \hat{H}_{SOC}=\alpha(\bold{\mathbf{s}\times k})\bold{\cdot}\bold{\hat{z}}.\label{eqn:formSOC}
\end{align} 
This is the only linear in wavevector $\bold{k}$=$(k_{x},k_{y})$ of electron (in the vicinity of the $\Gamma$-point) term allowed by $C_{3v}$ point-group symmetry of the lattice of $\gamma$-stacked multilayer (the next term in the $\bold{k\cdot p}$ theory expansion would be of the third order\footnote{The linear in wavevector SOC presbribed by the $C_{3v}$ point-group has the form $H^{(1)}_{SO}=i\alpha\Big(k_{-}\bold{s_{+}}-k_{+}\bold{s_{-}}\Big)$, where $k_{\pm}\equiv k_{x}\pm ik_{y}$ and $\bold{s_{\pm}}\equiv\frac{1}{2}(\bold{s_{x}}\pm i\bold{s_{y}})$. A higher order invariant in wavevector $k$ can be constructed replacing $k_{\pm}$ by $k^{3}_{\pm}$, therefore leading to a cubic SOC of the form $H^{(3)}_{SO}=-i\beta\Big(k^{3}_{-}\bold{s_{+}}-k^{3}_{+}\bold{s_{-}}\Big)=2\beta\Big((k^{3}_{x}-3k_{x}k^{2}_{y})\bold{s_{y}}-(3k_{y}k^{2}_{x}-k^{3}_{y})\bold{s_{x}}\Big)$.} in $k$, hence, much weaker for a feasible doping of the film)\cite{cubicRashba,winkler2003spin,Kochan}. In Eq. (\ref{eqn:formSOC}), $\bold{s}=(\sigma_{x},\sigma_{y})$ is a vector composed of Pauli  matrices, and $\alpha$ is a layer-number-dependent factor,
\begin{align}
    \alpha(\mathcal{E}_{z},N)\approx\alpha_{\infty}\Bigg(1-\frac{\chi}{(N+2.84)^{2}}\Bigg)\pm\mathcal{E}_{z}\aleph. \label{eqn:formSOCexp}
\end{align}
Here, $\alpha_{\infty}\approx34.5$ meV$\text{\AA}$ is the value of SOC at the conduction band edge of 3D bulk $\gamma$-InSe, $N$ is the number of layers in a thin film, $\chi\approx14.9$ accounts for the non-linear dependence of bulk SOC on the out-of-plane momentum $k_{z}$ counted from the bulk A-point band edge, at $k_{A}=\frac{\pi}{a_{z}}$.
Also $\mathcal{E}_{z}$ is the electric field piercing the film, and parameter $\aleph$ quantifies the dependence on the electric field, as shown in the inset of Fig. \ref{fig:RashbaEfield}.\\
\begin{table}[t!]
	\begin{tabular}{|c|c|c|c|c|c|}
		\hline\hline
		$E_{v} \quad$ & $-2.79$~eV & $ \quad \quad t^{\Gamma}_{cc} \quad \quad $ & $0.34$~eV & $m_{c}$ & $0.266~m_{0}$ \\
		$  t^{\Gamma}_{vv}$ & $-0.41$~eV & $ t_{cc_{2}}$& $-3.43$~eV\AA$^{2}$ & $E_{v_{1}}$&$-3.4$~eV\\
		$ E_{v_{2}}$&$-3.5$~eV & $t^{\Gamma}_{cv}$&$0.25$~eV &
		$ t_{cv_{2}}$&$-3.29$~eV\AA$^{2}$ \\
		\hline\hline
	\end{tabular}
\end{table}
\begin{table} [t!]
	\begin{tabular}{cccc}
		\hline\hline
		$ E_{cv} $& $ 2.79 $~eV & \\
	    $ E_{c_{1}c} $& $ 1.09 $~eV & \\
		$  E_{vv_{1}} $& $ 0.54 $~eV \\
		$  E_{vv_{2}} $& $ 0.683 $~eV \\
		$ b_{54}$& $ 10.54$~eV\AA & \\
		$ \lambda_{15} $& $ 0.119 $ ~eV& \\
		$ b_{16} $& $-2.77 $ ~eV\AA & \\
		$ b^{c_{1}v_{2}}_{16} $& $8.51 $ ~eV\AA & \\
		$ d_{cv} $& $-1.68$ e\AA & \\
		$ d_{v_{1}v_{2}}$& $-2.56$ e\AA & \\
	    $ d_{c_{1}c}$& $0.86$ e\AA & \\
	    $ t^{\Gamma}_{cc}$& $0.34$~eV \\
	    $ t^{\Gamma}_{vv}$ & $-0.41$~eV \\
		$ t^{\Gamma}_{cv}$ & $0.25$~eV & \\
		$ t_{cc_{1}}$ & $0.019$~eV & \\
		$ t_{v_{1}v_{2}}$ & $0.048$~eV & \\
		$\delta_{cv}$& 0.014~eV\\
		$\delta_{c_{1}c}$& 0.022~eV \\
		$\delta_{v_{1}v_{2}}$& -0.001~eV \\
		$\lambda_{46}$& -0.09~eV \\
		$a_{z}$ & 8.32 \AA \\
		\hline
	\end{tabular}\label{tbl:GWparameters14band}
	\caption{\footnotesize (Top) Two-band hybrid $\bold{k \cdot p}$ tight-binding parameters extracted from the 14-band model in the bottom table. (Bottom) Hybrid $\bold{k} \cdot \bold{p}$ tight-binding model parameters used in the perturbation theory analysis. Numerical indices in the $b$ and $\lambda$ terms label the symmetry group shown in the character table in Fig. \ref{fig:bandsandcharacter}. The magnitude of the out-of-plane dipole moments $d_{cv}$, $d_{v_{1}v_{2}}$ and $d_{c_{1}c}$ were obtained from the tight-binding model developed in Ref. \onlinecite{magorrian2016electronic}. The SOC parameter $\lambda_{46}$ was calculated from the fits performed in Appendix \ref{app:gammaepsilon} and the interlayer distance $a_{z}=8.32$\AA~ was obtained from the experimental measurements shown in Ref. \onlinecite{Rigoult1980}.}
\end{table}
\indent The overall strength of SOC in Eq. (\ref{eqn:formSOCexp}) is determined by the interplay between the intrinsic lattice asymmetry of the crystal (known as Dresselhaus contribution\cite{Dresselhaus}) and the electric-field-induced symmetry breaking (the so-called Bychkov-Rashba term\cite{Rashba}). This interplay allows for the tunability of the SOC value, both by choosing the film thickness ($Nd$), and by applying a displacement field in the double-gated (both top- and bottom-gated) devices.
The results of our analysis are exemplified in Fig. \ref{fig:RashbaEfield}, indicating that a vertically applied electric field $\mathcal{E}_{z}\sim$ 0.15-0.20 Vnm\textsuperscript{-1} would be enough to switch SOC off and on, opening new avenues towards the design of spintronic devices. This form of SOC in a film is the result of $\bold{k \cdot p}$ and tight-binding model analysis\cite{zhou2017multiband}, parameterized using density functional theory (DFT) computations of the band structure. The theoretically calculated SOC size was compared with the values of SOC strength extracted from weak antilocalization magnetoresistance, measured in a FET based on a six-layer InSe device. We find a good agreement between theory and experiment in the available range of device parameters. \newline
\indent Below, the paper is organized as follows. In Section \ref{section:DFT}, we compute the SOC coefficient in the lowest conduction subband of InSe using DFT \textit{ab intio} calculations, in Section \ref{section:bilayer}, we perform simple perturbative calculations of SOC strength in the lowest conduction subband of bilayer InSe and, in Section \ref{section:perturbationalaysis}, we generalise the bilayer formalism for an arbitrary number of layers. Finally, in Section \ref{section:magnetoconductivity}, we compare the theoretically obtained SOC coefficient with the values experimentally measured in an available InSe-based FET device.

\section{First principles calculations of I\MakeLowercase{n}S\MakeLowercase{e} parameters} \label{section:DFT}

\begin{figure}[b!]
    \centering
    \includegraphics[width = \linewidth]{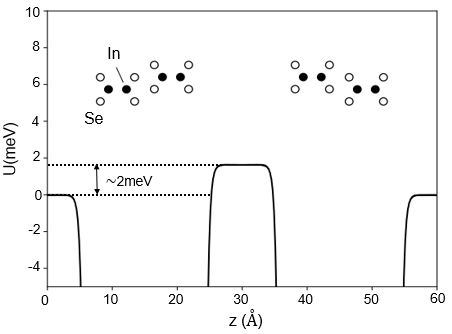}
    \caption{\footnotesize Plane-averaged electrostatic potential accounting for ionic and Hartree contributions in a double-bilayer InSe supercell (supercell structure shown as inset).}
    \label{fig:InSe_electrostatic_potential}
\end{figure}
As a background to the hybrid $\mathbf{k\cdot p}$ tight-binding (HkpTB) model presented in this manuscript, we overview the density functional theory bandstructure of monolayer and few-layer InSe.\newline
\begin{figure}[t!]
    \includegraphics[width=\columnwidth]{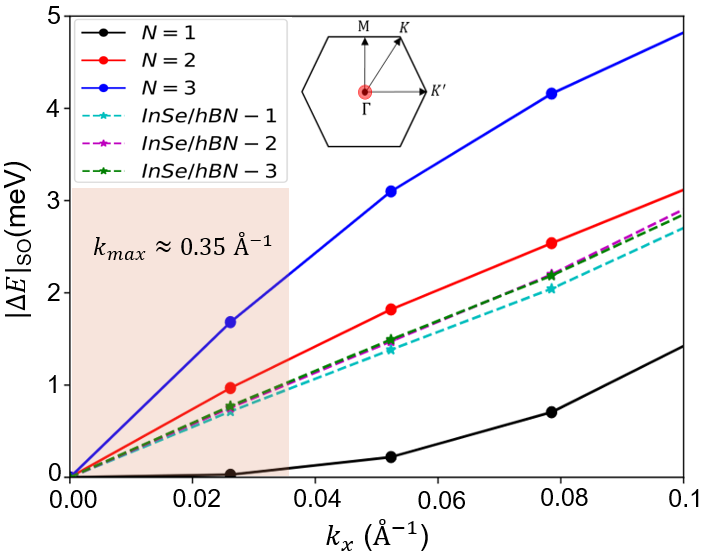}
    \caption{\footnotesize DFT-calculated conduction band spin-orbit splitting for mono-, bi-, and tri-layer InSe in for small $k_x$ near the $\Gamma$-point. The finite thickness of the film discretizes $k_{z}$ while $k_{x}$ and $k_{y}$ remain continuous variables. While the cubic Dresselhaus SOC splitting is expected to be zero in the $\Gamma-M$  direction, in the $(k_{x},0)$ orientation a finite contribution is expected. In contrast to the orientation-dependent cubic SOC, the expected form of the linear SOC splitting (see Eq. (\ref{eqn:formSOC})) makes this contribution isotropic in $\bold{k}$. The shaded region labels the range in $k_{x}$ below the Fermi level of a device doped with a carrier density of $n_{e}\approx2\times10^{12}$cm$^{-2}$. The clear linear spin splitting with $k_{x}$ indicates the dominance of the linear SOC terms near the Brillouin zone (BZ) center. Also plotted using stars connected by dashed lines are values of splitting for a monolayer InSe/monolayer hBN heterostructure for three different in-plane stacking configurations. (Inset) Hexagonal BZ of monolayer InSe. The red circle indicates the region in the BZ with wavevector magnitude in the range presented.}\label{fig:plainGW}
\end{figure}
Monolayer InSe has pairs of vertically aligned metal atoms in the middle sublayers and chalcogens in the outer sublayers, arranged on a plane into a honeycomb structure. Such a lattice has a $D_{3h}$ point-group symmetry which includes mirror plane symmetry, rotations by $120^{\circ}$, but not inversion symmetry. In any few-layer $\gamma$-InSe film, the $z\rightarrow -z$ mirror symmetry is broken. This opens a possibility for a weak ``ferroelectric'' charge transfer between the layers due to layer-asymmetric hybridization between the conduction and valence bands and the resulting built-in electric field in the film which may be relevant for the self-consistent analysis of the on-layer potential in a film with a finite thickness. To find out whether this is of relevance for InSe, or not, we carry out DFT calculations on a supercell with a large vacuum separating two mirror reflected images of a $\gamma$-InSe bilayer, to satisfy periodic boundary conditions without affecting the mismatch between vacuum potentials, produced by the double-charge layer due to the charge transfer
(see Fig. \ref{fig:InSe_electrostatic_potential} and inset).
For the DFT calculations, we used the generalized gradient approximation (GGA) of Perdew, Burke and Ernzerhof \cite{PBE}, with an $12\times 12 \times 1$ \textit{k}-point grid and a plane-wave cutoff energy of 600~eV, implemented in the VASP code\cite{VASP_PhysRevB.54.11169}.
Monolayer atomic structure parameters, and interlayer distances, are taken from an experimental reference for the bulk crystal\cite{Rigoult1980}. We find that the charge transfer between the layers is small, yielding a  $\approx$ 2 meV vacuum potential difference across the bilayer in Fig. \ref{fig:InSe_electrostatic_potential}, which is so small that it will be neglected for the rest of the manuscript.\newline
\indent Due to its mirror symmetry, the monolayer Hamiltonian cannot include $s_{x}$, and $s_{y}$ operators, that is, it does not display a 2D SOC. However, its symmetry allows for spin-orbit splitting in the form of\cite{li2015symmetry,Dresselhaus}
\begin{align}
    \hat{H}_{so}=\gamma k^{3}\sin(3\phi)\hat{s}_{z}\label{eqn:monospin}
\end{align}
where $\phi$ is the polar angle with respect to the $\Gamma-M$ direction and $\hat{s}_{z}$ is the third Pauli matrix. This is reflected by the results of DFT computations of conduction band dispersion in mono-, bi-, and trilayers shown in Fig. \ref{fig:plainGW}(a). 
\begin{table}[t!]
\scalebox{1}{
\begin{ruledtabular}
\begin{tabular}{c|cccccc|c|c|c|}
$D_{3h}$ & $E$ & $\sigma_{h}$ & $2C_{3}$ & $2S_{3}$ & $3C^{'}_{2i} $ & $3\sigma_{vi}$ &$basis$  & $orbitals $ & $ bands $\\
\hline
$A'_{1}(\Gamma_{1})$ & 1  & 1 &  1 &  1 & 1 & 1 & 1 & $(s^{+},p^{-}_{z})$ & $v,c_{1}$\\
$A'_{2}(\Gamma_{2})$ & 1  & 1 &  1 & 1 & -1 & -1 & $xy$ &  & \\
$E'(\Gamma_{6})$ & 2  & 2 &  -1 &  -1 & 0 & 0 & $(x,y)$ & $(p^{+}_{x},p^{+}_{y})$ & $v_{2}$\\
$A''_{1}(\Gamma_{3})$ & 1 & -1 & 1  & -1 & 1 & -1 & $xyz$ &  & \\
$A''_{2}(\Gamma_{4})$ & 1  & -1 & 1  &  -1 & -1 & 1 & $z$ & $(s^{-},p^{+}_{z})$ & $c$\\
$E''(\Gamma_{5})$ & 2  & -2 & -1  &  1 & 0 & 0 & $(xz,yz)$ & $(p^{-}_{x},p^{-}_{y})$ & $v_{1}$\\
\end{tabular}
\end{ruledtabular}}\label{tbl:characFer}
\end{table}
\begin{figure}[t!]
    \includegraphics[width=\columnwidth, height=0.8\columnwidth]{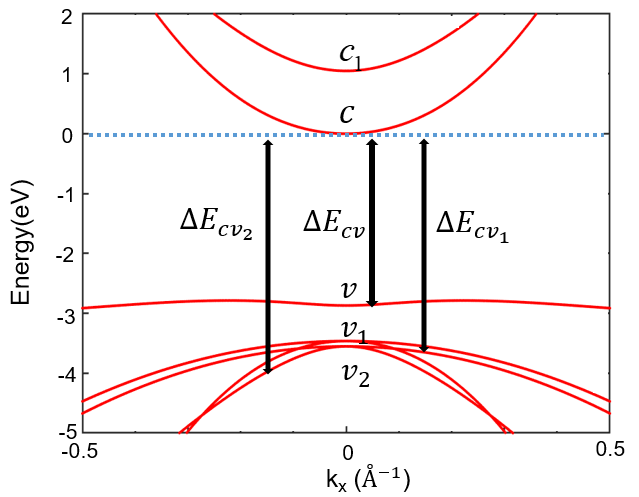}
    \caption{\footnotesize (Top) Character table of the point-group $D_{3h}$ which captures the symmetries of monolayer InSe. In parenthesis, the Bethe notation for each irrep is shown. Both the basis function of each irreducible representation as well as the orbital composition of any band relevant for our analysis are displayed in the final columns. The $\sigma_{h}$ conjugacy class in the character table labels the $z\rightarrow-z$ symmetry of each irreducible representation. This crucially determines which bands are mixed due to an applied electric field. The superscripts on top of the orbitals indicate the parity with respect to the $z\rightarrow-z$ symmetry calculated in Ref. \onlinecite{li2015symmetry}. (Bottom) Band structure of monolayer InSe without SOC.}\label{fig:bandsandcharacter}
\end{figure}Note that the spin polarization of the computed states is in $z$-direction only for monolayers, whereas for bi- and trilayers, where it has a linear dependence announced in Eq. (\ref{eqn:formSOC}), it reflects in-plane spin splitting. In fact, for the range of in-plane wavenumbers corresponding to feasible doping densities, the spin splitting in the monolayer is negligibly small.  \footnote{Note that the SOC coefficients, $\alpha$, implied by the DFT results, 18.6~meV\AA~and 32.3~meV\AA~for bilayer and trilayer respectively, are somewhat larger than those predicted by the model presented in this work - this is due to the substantial underestimation of the band gap of InSe by DFT. We therefore base the parametrization of the model on GW results for bulk crystals}. We also carried out DFT calculations for a heterobilayer consisting of monolayer of InSe, and monolayer of hBN (the latter was strained to give commensurability with a lattice constant $a_{\mathrm{hBN}} = a_{\mathrm{InSe}}/\sqrt{3}$ and rotated to align the armchair direction of hBN with the zigzag direction of the InSe). We take the interlayer distance as 0.333~nm between the middle of hBN and the nearest plane of Se atoms. A dipole correction was applied, and we considered three in-plane configurations: (1) boron directly above indium, (2) nitrogen above indium (the hBN is inverted in-plane), and (3) configuration (1) with the hBN shifted in-plane by half the B-N vector. The spin-orbit splitting near $\Gamma$ in the (InSe-dominated) conduction band edge is plotted for all 3 configurations in Fig. \ref{fig:plainGW}. For the monolayer InSe/monolayer hBN heterostructure, we obtain a SOC which depends very weakly on the configuration, with a strength similar to that of the isolated InSe bilayer.
\begin{figure}[t!]
   \includegraphics[width=\columnwidth]{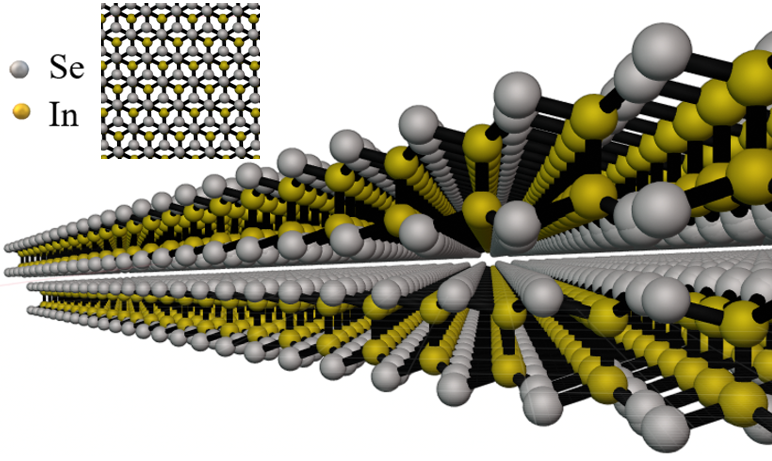}
    \caption{\footnotesize Profile and top view of bilayer $\gamma$-stacked InSe. The Se atom of the top layer is shown to sit above the In atom of the bottom layer but not the other way around. This crystallographic \textit{z}-asymmetry is responsible for an effective ``electric field'' at the origin of the Dresselhaus SOC in bilayer InSe.}\label{fig:bilayerstack}
\end{figure}
\section{Spin-orbit coupling in I\MakeLowercase{n}S\MakeLowercase{e} bilayer }\label{section:bilayer}

InSe belongs to the family of group-III metal-monochalcogenides with the $s$ and $p$ orbitals of In and Se dominating the low-energy dispersion in the vicinity of the $\Gamma$-point\cite{magorrian2016electronic,li2015symmetry,zhou2017multiband}. In the bottomost conduction band $c$ and in the topmost valence band $v$, the atomic orbital composition is mainly dominated by the $s$ and $p_{z}$ orbitals of both In and Se. The deeper valence bands $v_{1}$ and $v_{2}$ are prominently Se $p_{x}$ and $p_{y}$ orbitals which are naturally split by the atomic SOC of the Se atoms. \newline 
\indent In going from monolayer to bilayer $\gamma$-stacked InSe (see Fig. \ref{fig:bilayerstack}), the mirror plane symmetry is broken, reducing the symmetry from the point-group $D_{3h}$ to $C_{3v}$. This allows for a linear in momentum SOC splitting in the form presented in Eq. (\ref{eqn:formSOC}) prescribed by the third-order rotation symmetry axis\cite{SOCgraphite}. Consequently, the interlayer hoppings need to account for the reduction of the global symmetries of the bilayer, leading to a finite Dresselhaus SOC. This appears via the interlayer mixing of the opposite $z$-parity bands. \newline
\indent We construct a bilayer Hamiltonian using monolayer Hamiltonians described in Ref. \onlinecite{magorrian2016electronic} taking into account interlayer hopping\cite{magorrian2016electronic} and the intralayer interband spin-orbit coupling\cite{soc2017},
\begin{align}
    \hat{H}=\hat{H}^{(0)}+\delta\hat{H}=\begin{pmatrix}
        \hat{H}_{11}^{(0)}+\delta \hat{H}_{11}&\hat{H}_{12}^{(0)}+\delta \hat{H}_{12}& \\
        \hat{H}_{21}^{(0)}+\delta \hat{H}_{21}&\hat{H}_{22}^{(0)}+\delta \hat{H}_{22}&\\
    \end{pmatrix},\label{eqn:basicbi}
\end{align}
For the analysis of SOC in the bilayer, band edge states in the constituent monolayers, 
\begin{align*}
\Psi^{\text{T}}\equiv[&c^{\uparrow},c^{\downarrow};c^{\uparrow}_{1},c^{\downarrow}_{1};v^{\uparrow},v^{\downarrow};v^{\uparrow,p_{x}}_{1},v^{\downarrow,p_{x}}_{1};v^{\uparrow,p_{y}}_{1},v^{\downarrow,p_{y}}_{1};\nonumber \\
&
v^{\uparrow,p_{x}}_{2},v^{\downarrow,p_{x}}_{2};v^{\uparrow,p_{y}}_{2},v^{\downarrow,p_{y}}_{2}], \\ \nonumber
\end{align*}
for the bands described in Section \ref{section:DFT}, will be characterised by their respective band energies, neglecting an almost parabolic band dispersion, 
\begin{widetext}
\begin{align}
    \hat{H}_{11(22)}^{(0)}=
    \begin{pmatrix}
        -U_{1(2)}&0&0&0&0& \\
        0&E_{c_{1}}-U_{1(2)}&0&0&0& \\
        0&0&E_{v}-U_{1(2)}&0&0& \\
        0&0&0&(E_{v_{1}}-U_{1(2)})\bold{\hat{I}}_{\nu}&0&\\
        0&0&0&0&(E_{v_{2}}-U_{1(2)})\bold{\hat{I}}_{\nu}&\\
    \end{pmatrix}.\label{eqn:unpertmono}
\end{align}
\end{widetext}
Here $\bold{\hat{I}}_{\nu}$ is the identity operator in the $2\times2$ space of atomic $p_{x}, p_{y}$ orbital components of $v_{1}$ and $v_{2}$. $E_{c_{1}},E_{v},E_{v_{1}}$ and $E_{v_{2}}$ are the energy differences between the lowest conduction band and the $c_{1},v,v_{1}$ and $v_{2}$ bands, respectively, (see Fig. \ref{fig:bandsandcharacter}). In addition, we take into account linear in momentum interband terms in the monolayer Hamiltonian , discussed earlier in relation to the optical selection rules for the interband transitions\cite{soc2017},
\begin{widetext}
\begin{align}
    \centering
    \delta \hat{H}_{11(22)}=
    \begin{pmatrix}
        0&0&0&ib_{54}\bold{k} \cdot \bold{\Lambda}&i\lambda_{46}(\bold{s}\times \bold{\Lambda})&\\
        0&0&0&0&ib^{c_{1}v_{2}}_{16}(\bold{k} \cdot \bold{\Lambda})&\\
        0&0&0&i\lambda_{15}(\bold{s}\times \bold{\Lambda})&ib_{16}(\bold{k} \cdot \bold{\Lambda})&\\
        -ib_{54}(\bold{k} \cdot \bold{\Lambda})^{T}&0&-i\lambda_{15}(\bold{s}\times \bold{\Lambda})^{\dag}&0&0&\\
        -i\lambda_{46}(\bold{s}\times \bold{\Lambda})^{\dag}&-ib^{c_{1}v_{2}}_{16}(\bold{k} \cdot \bold{\Lambda})^{T}&-ib_{16}(\bold{k} \cdot \bold{\Lambda})^{T}&0&0& \\ 
    \end{pmatrix},\label{eqn:pertmono}
    \end{align}
\end{widetext}
Here $1\times2$ matrices $\bold{\Lambda_{y}}$=[0,1] and $\bold{\Lambda_{x}}$=[1,0] operate in the $p_{x}, p_{y}$ orbital components of $v_{1}$ and $v_{2}$ valence bands and the coefficients $b_{45}, b_{16}$ and $b^{c_{1}v_{2}}_{16}$ characterise the $c-v_{1}$, $v-v_{2}$ and $c_{1}-v_{2}$ intra-layer couplings (associated with interband optical transitions excited by the in-plane polarised photons). Spin Pauli matrices $\bold{s_{x,y}}$ produce spin flips upon the interband mixing which can be rooted to atomic $\bold{S\cdot L}$ coupling (between $p_{x/y}$ and $p_{z}$ orbitals which contribute to $c,v,v_{1},v_{2}$ bands captured by parameters $\lambda_{15}$ and $\lambda_{46}$). Note that $\bold{k} \cdot \bold{\Lambda}\equiv k_{x}\bold{\Lambda_{x}}+k_{y}\bold{\Lambda_{y}}$ and $\bold{s}\times \bold{\Lambda}\equiv \bold{s_{x}}\bold{\Lambda_{y}}-\bold{s_{y}}\bold{\Lambda_{x}}$.\\
\indent Hopping between neighbouring layers is accounted for by the following two terms,
\begin{widetext}
    \begin{align}
    \hat{H}_{12}^{(0)}=
    \begin{pmatrix}
        t^{\Gamma}_{cc}&0&0&0&0 \\
        0&0&0&0&0 \\
        0&0&t^{\Gamma}_{vv}&0&0 \\
        0&0&0&0&0\\
        0&0&0&0&0\\
    \end{pmatrix},
    \end{align}
    \begin{align}
    \delta \hat{H}_{12}=\begin{pmatrix}
        0&(t_{cc_{1}}+\delta_{c_{1}c})&(t^{\Gamma}_{cv}+\delta_{cv})&0&0\\
        (-t_{cc_{1}}+\delta_{c_{1}c})&0&0&0&0\\
        (-t^{\Gamma}_{cv}+\delta_{cv})&0&0&0&0 \\
        0&0&0&0&0 \\
        0&0&0&0&0 \\
    \end{pmatrix}.\label{eqn:deltaHbilayer}
\end{align}
\end{widetext}
The first of them describes the resonant interlayer hybridization of separately lower conduction and the top valence band edges, which was identified \cite{magorrian2016electronic} as the strongest hybridization effect, determined by the substantial weight of $s$ and $p_{z}$ chalcogen orbitals in the sublattice composition of the band edge states. The second term takes into account interband interlayer hybridization, which produces a much weaker effect on the band edge energies, but is sensitive to the mirror symmetry breaking set by stacking of the layers (see Fig. \ref{fig:bandsandcharacter}).\\
\indent According to the table in Fig. \ref{fig:bandsandcharacter}, the on-layer states in bands $c$ are odd under $z\rightarrow-z$ reflection while bands $v$ and $c_{1}$ are even under the same transformation. Because of this, for a mirror symmetric arrangement of the layers, the corresponding interband interlayer couplings would obey the relation $t^{\Gamma}_{cv}=-t^{\Gamma}_{vc}$ and $t_{c_{1}c}=-t_{cc_{1}}$. To capture the mirror plane symmetry breaking for $\gamma$-stacking, we introduce parameters $\delta_{\alpha\beta}$ such that $t^{\Gamma}_{cv}=t^{\Gamma}_{cv}+\delta_{cv}$, $t^{\Gamma}_{vc}=-t^{\Gamma}_{cv}+\delta_{cv}$, $t_{v_{1}v_{2}}=t_{v_{1}v_{2}}+\delta_{v_{1}v_{2}}$, $t_{v_{2}v_{1}}= -t_{v_{1}v_{2}}+\delta_{v_{1}v_{2}}$, $t_{cc_{1}}=t_{cc_{1}}+\delta_{c_{1}c}$, and $t_{c_{1}c}=-t_{cc_{1}}+\delta_{c_{1}c}$. Overall, the $z\rightarrow-z$ symmetry breaking in the bilayer (which gives rise to the 2D SOC in the lowest conduction subband of the bilayer) is produced by the interplay between $\delta H_{11}$ and the contributions from $\delta_{\alpha\beta}$ in Eq. (\ref{eqn:deltabilayer}). For this we use $3^{rd}$ order pertubation theory with respect to parameters $\delta_{c_{1}c},\delta_{cv},b_{54},b_{16},b^{c_{1}v_{2}}_{16},\lambda_{15}$, and $\lambda_{46}$, and this results in the spin-orbit coupling constant,
\begin{align}
   \alpha_{0}=2\Bigg( \frac{b_{54}\lambda_{15}\delta_{cv}}{\Delta E_{cv_{1}}\Delta E_{g^{1}}}+\frac{b_{16}\lambda_{46}\delta_{cv}}{\Delta E_{cv_{2}}\Delta E_{g^{1}}}+\frac{b^{c_{1}v_{2}}_{16}\lambda_{46}\delta_{c_{1}c}}{\Delta E_{cc_{1}}\Delta E_{cv_{2}}}\Bigg).\label{eqn:deltabilayer}
\end{align}
Here we also account for asymmetry induced by an external electric field so its effect on the on-layer energy of the orbitals in Eq. (\ref{eqn:unpertmono}), captured by $\Delta E_{cv_{1}}\equiv -(t^{\Gamma}_{cc}+E_{v_{1}})$, $\Delta E_{cv_{2}}\equiv-(t^{\Gamma}_{cc}+E_{v_{2}})$ and $\Delta E_{cc_{1}}\equiv -(t^{\Gamma}_{cc}+E_{c_{1}})$ are the energy differences between the lowest conduction subband and $v_{1}$, $v_{2}$ and $c_{1}$ bands while $\Delta E_{g^{1(2)}}=-(t^{\Gamma}_{cc}+E_{v})\pm t^{\Gamma}_{vv}$ is the energy difference between the lowest conduction subband and the $1^{st}$ or $2^{nd}$ topmost valence subband, respectively.\\
\indent In the absence of external electric field, $U_{1}=U_{2}=0$, and using parameters in Table \ref{tab:ml_kp_parameters}, we estimate that $\mathcal{E}_{z}(\alpha_{0}=0)=0.35$Vnm$^{-1}$. The dependence on a perpendicularly applied electric field $\mathcal{E}_{z}$ is approximated by
\begin{align}
    &\label{eqn:slopebilayer}
    \aleph\equiv\frac{d\alpha}{d\mathcal{E}_{z}}\bigg|_{U_{1}=U_{2}=0}=\frac{(b_{54}\lambda_{15}+b_{16}\lambda_{46})ea_{z}t^{\Gamma}_{cv}}{\Delta E_{cv_{1}}}\\ \nonumber
    &
    \times\Bigg(\frac{2t^{\Gamma}_{vv}}{\Delta E_{g^{1}}\Delta E_{g^{2}}}\Bigg)\Bigg(\frac{1}{2 t^{\Gamma}_{cc}}-\frac{1}{2t^{\Gamma}_{vv}}\Bigg).
\end{align}
Here, $a_{z}=8.32$~\AA~ is the interlayer distance between the central planes of two neighbouring InSe monolayers. Using parameters in Table \ref{tbl:GWparameters14band} we estimate that for a bilayer $\aleph=38$meV\AA/Vnm$^{-1}$,
this also means that an electric field $\mathcal{E}_{z}=0.35$~Vnm\textsuperscript{-1} would reduce the 2D SOC coupling strength to zero.\\
\indent In addition to the above-discussed effects, mirror symmetry breaking may be caused by the encapsulation environment\cite{dipoleinterface} coupling on the Se orbitals in the outer top/bottom sublayers of the crystal. This asymmetry may be due to the difference between the encapsulating materials, or even due to a different orientation of the top/bottom encapsulating layers of the same compound, e.g., hexagonal boron nitride (hBN). To describe this effect, we introduce an additional term in the bilayer Hamiltonian responsible for $c-v$, $v_{1}-v_{2}$ band mixing with randomly different strength in the top and bottom layers,
\begin{widetext}
\begin{align}
    \delta\hat{H}^{(I)}_{11(22)}=\begin{pmatrix}
        \Delta E_{c1(2)}&0&\pm\Upsilon^{t/b}_{cv}&0&0& \\
        0&0&0&0&0& \\
        \pm\Upsilon^{t/b}_{cv}&0&\Delta E_{v1(2)}&0&0& \\
        0&0&0&\Delta E_{v_{1}1(2)}\bold{\hat{I}}_{\nu}&\pm\Upsilon^{t/b}_{v_{1}v_{2}}\bold{\hat{I}}_{\nu}&\\
        0&0&0&\pm\Upsilon^{t/b}_{v_{1}v_{2}}\bold{\hat{I}}_{\nu}&\Delta E_{v_{2}1(2)}\bold{\hat{I}}_{\nu}&\\ \label{eqn:H0bilinter}
    \end{pmatrix}.\\ \nonumber
\end{align}
\end{widetext}
Here, $\Delta E_{c1(2)}$ and $\Delta E_{v1(2)}$ are the energy shifts of the $c$ and the $v$ bands in the $1^{st}$ and $2^{nd}$ layer respectively; $\Delta E_{v_{1}1(2)}$ and $\Delta E_{v_{2}1(2)}$ are the energy shifts of the bands $v_{1}$ and $v_{2}$ and $\bold{\hat{I}}_{\nu}$ is the identity operator in the $2\times2$ space of atomic $p_{x}, p_{y}$ orbital components of the $v_{1}$ and $v_{2}$ bands. The terms $\Upsilon^{t}_{cv}$ and $\Upsilon^{t}_{v_{1}v_{2}}$ are responsible for $c-v$ and $v_{1}-v_{2}$ band mixing in the top layer: the interfacial $z\rightarrow-z$ symmetry breaking couples states of opposite parities. In the bottom surface, the interfacial effect is inverted, which is the reason for the inverted signs, $-\Upsilon^{b}_{cv}$ and $-\Upsilon^{b}_{v_{1}v_{2}}$ of the corresponding terms in $\delta\hat{H}^{(I)}_{11(22)}$.
\begin{table}[b!]
    \centering
    \begin{tabular}{c|cccc}
    \hline\hline\hline
    InSe/hBN stacking & $\Delta E_c$ & $\Delta E_v$ & $\vert\Upsilon_{cv}\vert$ & $\vert\Upsilon_{v_{1}v_{2}}\vert$ \\ 
  \hline 1  & 140meV & 141meV &  35.6meV & 36.98meV  \\
         2 &  155meV & 95meV &  20.5meV & 32.77meV \\
         3 &  146meV & 141meV &  35.6 meV & 39.37meV \\
        \hline
    \end{tabular}
    \caption{DFT-estimated parameters describing the effect of hBN substrate or overlay on an InSe monolayer in Eq. (\ref{eqn:H0bilinter}).}\label{tab:InSe_hBN_shifts}
\end{table} In Table \ref{tab:InSe_hBN_shifts}, we quote values of all those parameters obtained using DFT modelling described in Section \ref{section:DFT}. In order to extract those parameters, the wavefunctions of bands $c$ and $v_{1}$ were obtained for the three different atomic arrangements described in Section \ref{section:DFT}. By comparing their wavefunction distribution with the DFT-computed wavefunctions of suspended monolayer InSe, the mixing terms between opposite $z$-parity bands $\Upsilon_{cv}$ and $\Upsilon_{v_{1}v_{2}}$  was extracted for each configuration. Finally, from the DFT energy eigenvalues, the shifts in energy of bands $c$ and $v$ were obtained for each of the three different configurations; the energy shifts of bands $v_{1}$ and $v_{2}$ were neglected due to the very weak interlayer hybridization of those bands which results in a negligible contribution to the conduction band SOC strength. Using pertubation theory, we calculate the contribution of these additional terms towards bilayer SOC and find that the dominant effect comes from the $c-v$ band mixing, resulting in,
\begin{align}
    &
    \alpha^{(I)}=\Big[\frac{b_{45}\lambda_{15}}{\Delta E_{g^{1}}\Delta E_{cv_{1}}}+\frac{b_{16}\lambda_{46}}{\Delta E_{g^{1}}\Delta E_{cv_{2}}}\Big]\Big(\Upsilon^{t}_{cv}-\Upsilon^{b}_{cv}\Big).\\ \nonumber
\end{align}
The above equation suggests that encapsulation of InSe with the same material in the top and bottom would result in the cancellation of the main part of such an additional contribution. Due to misalignement or an offset of the encapsulating crystals, this cancelation would never be exact leaving a residual effect due to the variation of InSe and, e.g., hBN stacking. Taking into account the random nature of such a variation, in the mechanically assembled structures, we estimate characteristic size of the residual SOC contribution using the characteristic difference of the $\Upsilon_{cv}$ parameters for two InSe/hBN stackings analysed in Section \ref{section:DFT} (Configuration 1 and 2 in Table \ref{tab:InSe_hBN_shifts} and Fig. \ref{fig:plainGW}). This gives $\vert\alpha^{(I)}\vert\sim3.5$meV\AA, which is an order of magnitude smaller than $\alpha_{0}=13$meV\AA. As a result, for InSe bilayer encapsulated with hBN on both sides, the value and displacement field dependence of SOC can be well described using Eq. (\ref{eqn:deltabilayer}) and (\ref{eqn:slopebilayer}).

\section{Spin-orbit coupling in multilayer I\MakeLowercase{n}S\MakeLowercase{e}}\label{section:perturbationalaysis}

Here, we combine the analysis of two factors that determine the strength of SOC in multilayer $\gamma$-InSe: the asymmetry embedded into the interlayer hybridization and the effect of an externally controllable electric field.

\subsection{Self-consistent analysis of subband electrostatics in doped multilayer I\MakeLowercase{n}S\MakeLowercase{e} films}\label{section:selfconsistentsec}

In this section, the effect of an externally applied electrostatic potential (gating) for electrons in the lowest conduction subband is calculated self-consistently, and its effect on the charge distribution and on the band gap is discussed for the dual and single-gated FET geometry as sketched in insets of Fig. \ref{fig:carrierdisp} and Fig. \ref{fig:mainresult},
respectively. To quantify the SOC in the lowest conduction subband of few-layer InSe films, we describe the subband structure of the latter (both dispersion and wavefunctions) taking into account the electrostatic potential profile induced by doping and gating. Our `workhorse' is a 2-band hybrid $\mathbf{k\cdot p}$ tight-binding (HkpTB) model previously discussed in Ref. \onlinecite{subbands2018}, formulated in the basis of conduction, $c$ and valence, $v$ band states in each layer $[c_{1},v_{1},c_{2},v_{2},...]$. The HkpTB Hamiltonian has the form,
\begin{align}
    \hat{H}^{N}_{\mathbf{k\cdot p}} \approx
    \begin{pmatrix}
        \frac{\hbar^{2}k^{2}}{2m_{c}}+U_{1}&0&t_{cc}&t_{cv}&\cdots\\\ 0&E_{v}+U_{1}&-t_{cv}&t_{vv}&\cdots \\ t_{cc}&-t_{cv}&\frac{\hbar^{2}k^{2}}{2m_{c}}+U_{2}&0&\cdots\\ t_{cv}&t_{vv}&0&E_{v}+U_{2}&\cdots\\
                        0&0&t_{cc}&0& \cdots \\
                        \vdots&\vdots&\vdots&\vdots&
        \end{pmatrix}\label{eqn:2band}
\end{align}
Here, $t_{cc(vv)}$ parameterize the interlayer conduction-conduction (valence-valence) hops ($t_{cc}\equiv t^{\Gamma}_{cc}+t_{cc_{2}}k^{2}$), while $t_{cv}$ ($t_{cv}\equiv t^{\Gamma}_{cv}+t_{cv_{2}}k^{2}$) is the conduction to valence band hop. The zero of energy is set to the monolayer conduction band edge, so that $E_{v} \approx-2.8$~eV is the energy of the monolayer's topmost valence band at the $\Gamma$-point. We neglect the valence band dispersion in InSe monolayers, as earlier studies\cite{hammerindirect,rybkovskiy_transition,Katsnelson_Phonons,Zolyomi2014} have shown that it is approximately flat over a large central part of the Brillouin zone. We also neglect any $k$-dependence in $t_{vv}$ for the same reason. The terms $U_{\eta}$ account for the electrostatic potential in layer $\eta$, and they are calculated as\cite{subbands2018},
\begin{widetext}
\begin{figure*}[!t]
    \includegraphics[width=7.2in,height=2.3in]{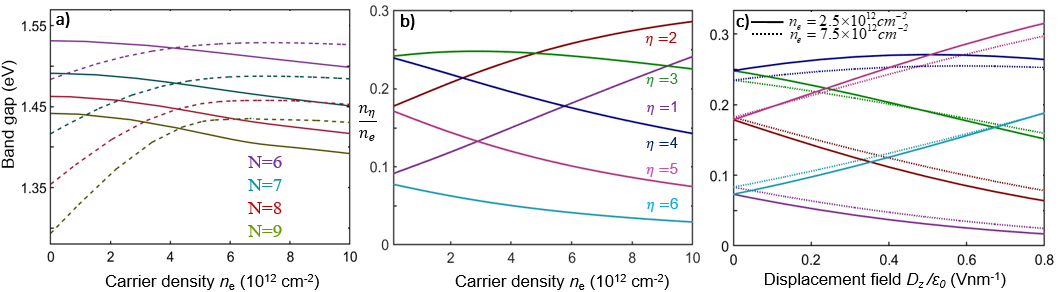}
	\caption{\footnotesize (a) Band gap dependence on carrier density for a single (solid) and dual-gated (dashed) device with a fixed top gate carrier density $n_{tg}=4\times 10^{12}$cm\textsuperscript{-2}. A reduction in the band gap with increasing electric field is expected from the displacement of electrons towards lower energies along with an increase in electrostatic energy of the holes (quantum-confined Stark effect)\cite{tunelJohanna,QCSE}. (b) Fraction of the total carrier density at each layer $\eta$ in a single-gated six-layer InSe device. The first layer is defined as the one closest to the metallic gate. (c) Fraction of the total  carrier density at each layer $\eta$ against displacement field in a six-layer InSe device in a dual-gated configuration at two fixed carrier concentrations of $n_{e}=2.5\times10^{12}$cm\textsuperscript{-2} for the solid line and $n_{e}=7.5\times10^{12}$cm\textsuperscript{-2} for the dotted line. The same color to layer correspondence applies as in Fig. 2(b).}\label{fig:carrierlayer}
\end{figure*}
\end{widetext}
\begin{equation}
U_{\eta>1}=U_{1}+ea_{z}\sum_{\kappa=2}^{\kappa=\eta}\mathcal{E}_{(\kappa-1)\kappa},
\end{equation}
where $a_z=8.32$~\AA~is the distance between adjacent layers and $\mathcal{E}_{(\kappa-1)\kappa}$ is the electric field between layers $\kappa-1$ and $\kappa$. $\mathcal{E}_{(\kappa-1)\kappa}$ is obtained from the electron density on each InSe layer, $n_{\eta}$, as
\begin{equation}
\mathcal{E}_{(\kappa-1)\kappa}=\frac{e}{\epsilon_{z}\varepsilon_0}\sum_{\eta=\kappa}^{\eta=N}n_{\eta},\label{eqn:sce}
\end{equation}
where $N$ is the total number of InSe layers in the device, $n_{\eta}$ is the carrier concentration at the $\eta^{th}$ layer and $\epsilon_{z}$ is the dielectric constant of InSe in the \textit{z}-direction. We then approximate the electric field across a single layer as the mean of the fields either side of it,
\begin{equation}
\mathcal{E}_{\kappa} \simeq (\mathcal{E}_{(\kappa-1)\kappa}+\mathcal{E}_{\kappa(\kappa+1)})/2.\label{eqn:averageE}
\end{equation}
Values of the parameters in the above Hamiltonian are listed in Table \ref{tab:ml_kp_parameters}. They are obtained by fitting the results of the numerical analysis of the 14-band model described in Ref. \onlinecite{magorrian2016electronic,subbands2018,ExcitonsGW,Questaal}.
It is also common, in order to obtain more flexibility in gating, to have both a back gate and a top gate applied to the device as shown in the dual-gated geometry in the inset of Fig. \ref{fig:carrierdisp}. To demonstrate the behaviour of the SOC coefficient in the dual-gated case, we reproduce the gating configuration used for transport experiments on a six-layer device studied in Ref. \onlinecite{bandurin2017high}. In that work, a fixed positive top gate voltage was applied to dope the system. At $V_{bg}=0$, the carrier density in the InSe films was measured to be $n_{e}\sim 4\times 10^{12}$ cm\textsuperscript{-2} indicating that the charge density in the top plate was that same amount. To include a fixed top gate in our electrostatic calculations, we amend Eq. (\ref{eqn:sce}) to read
\begin{equation}
    \mathcal{E}_{(\kappa-1)\kappa}=\frac{e}{\epsilon_{z}\varepsilon_0}\left[\sum_{\eta=\kappa}^{\eta=N}n_{\eta}-n_{tg}\right],
\end{equation}
where $n_{tg}$ is the fixed top gate carrier density and $n_{\eta}$ the carrier density in layer $\eta$. In considering the single-gated FET geometry, a band gap modulation in the range of 10$-$20~meV is obtained for carrier densities in the range of $0-3\times 10^{12}$cm\textsuperscript{-2} for 6$-$9 layers as shown in Fig. \ref{fig:carrierlayer}(a). Such band gap tunability\cite{bandurin2017high} is a lot more efficient in the dual-gated configuration, due to a reduced electrostatic screening, with the band gap increasing up to 50 meV for an 8 layer device with a doping density of $2\times 10^{12}$ cm\textsuperscript{-2} and an applied top gate carrier density of $n_{tg}=4\times 10^{12}$ cm\textsuperscript{-2}. This reduction in screening also makes the charge redistribution more efficient in the dual-gated FET device compared with the single-gated configuration, see  Fig. \ref{fig:carrierlayer}(b) and \ref{fig:carrierlayer}(c).
\begin{table}[H]
\centering
\begin{tabular}{|c|c|c|}
\hline
\centering
$ L $ & $\quad Band\quad gap$ (eV) & $m_c/m_0$ \\
\hline
1 & \quad \quad 2.87  \quad \quad & \quad \quad 0.266 \quad \quad\\
2 & \quad \quad 2.14 \quad \quad  & \quad \quad 0.220 \quad \quad\\
3 & \quad \quad 1.83 \quad \quad  & \quad \quad 0.204 \quad \quad\\
4 & \quad \quad 1.67 \quad \quad  & \quad \quad 0.197 \quad \quad\\
5 & \quad \quad 1.58 \quad \quad  & \quad \quad 0.192 \quad \quad\\
6 & \quad \quad 1.52 \quad \quad  & \quad \quad 0.189 \quad \quad\\
7 & \quad \quad 1.48  \quad \quad & \quad \quad 0.187 \quad \quad\\
8 & \quad \quad 1.46 \quad \quad  & \quad \quad 0.186 \quad \quad\\
9 & \quad \quad 1.44  \quad \quad & \quad \quad 0.185 \quad \quad\\
10& \quad \quad 1.42  \quad \quad & \quad \quad 0.184 \quad \quad\\
\hline
\end{tabular}
\caption{\footnotesize Dependence of the energy gap and of  the effective mass of the lowest conduction subband as a function of the number of layers $L$; $m_{0}$ is the free electron mass.}\label{tab:ml_kp_parameters}
\end{table}

\subsection{SOC in multilayer films from few-layer HkpTB}

\begin{figure}[t!]
    \centering
    \includegraphics[width=\columnwidth]{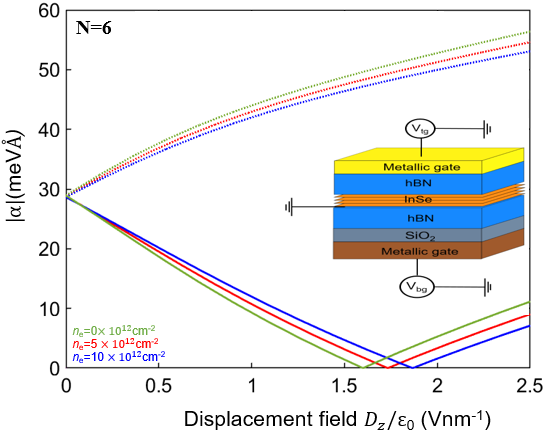}
	\caption{\footnotesize SOC strength dependence on displacement field and carrier density for a six-layer InSe dual-gated FET device as shown in the inset. The solid and dotted lines indicate respectively when the applied displacement field suppresses or enhances the Dresselhaus SOC.}
	\label{fig:carrierdisp} 
\end{figure}

\begin{figure}[t!]
    \centering
    \includegraphics[width=\columnwidth,height=4in]{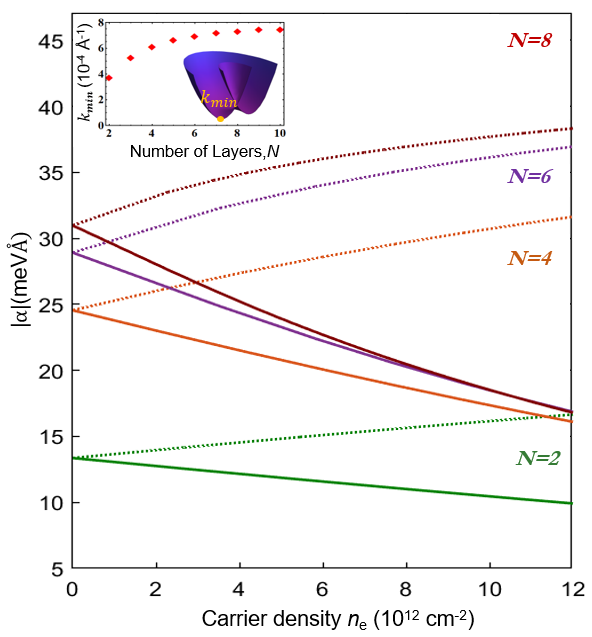}
	\caption{\footnotesize 2D SOC coefficient $\alpha$ in the lowest conduction subband of N-layer InSe film against carrier density for different number layers in a single-gated FET geometry. Inset: shift in momentum of the minimum of the lowest conduction subband as a function of the number of layers when no electrostatic doping is present. The dielectric constant, used for this calculation, was $\epsilon_{z}$=9.9. The solid and dotted lines indicate respectively when the applied displacement field suppresses or enhances the Dresselhaus SOC.}\label{fig:Rashbacarrier}
\end{figure}

In analyzing the SOC in multilayer InSe, two main mechanisms are found to determine the SOC strength. First, there are the intralayer dipole moments which mix wavefunctions of opposite parities within each layer under an applied electric field. Second, there is an interplay between the intrinsic inversion asymmetry of the lattice structure of $\gamma$-InSe, and the overall wavefunction $z \rightarrow -z$ symmetry breaking due to the applied electrostatic potential.
For the analysis of SOC in multilayer InSe it is necessary to include deeper valence bands $v_{1}$ and $v_{2}$ dominated by the $p_{x},p_{y}$ orbitals necessary for atomic SOC mixing with the $p_{z}$ orbitals in $c$ and $v$ (see the orbital composition of each band in the character table on top of Fig. \ref{fig:bandsandcharacter}). 
On including the deeper valence bands, the hybrid $\bold{k \cdot p}$ tight-binding Hamiltonian $\hat{H}$ of an $N$-layer InSe\cite{zhou2017multiband} in the vicinity of the $\Gamma$-point ($k_{x},k_{y}\rightarrow 0$) previously discussed in Section \ref{section:bilayer} is  rewritten as the sum of an unperturbed $\hat{H}^{(0)}$ and a perturbative part $\delta \hat{H}$,
\begin{align}
    \hat{H}=\hat{H}^{(0)}+\delta\hat{H}.
\end{align}
Writing the wavefunction eigenstates of the multilayer Hamiltonian $\hat{H}$ in a $14\times N$ band basis as $\Psi=[\Phi_{1},\Phi_{2},\Phi_{3},\Phi_{4},...,\Phi_{N}]$, where $\Phi_{w}$ is the 14-band monolayer basis in layer $w$ defined as 
\begin{align}
\Phi_{w}\equiv[&c^{\uparrow(w)},c^{\downarrow(w)},c_{1}^{\uparrow(w)},c_{1}^{\downarrow(w)},v^{\uparrow(w)},v^{\downarrow(w)},v^{\uparrow,p_{x}(w)}_{1},v^{\downarrow,p_{x}(w)}_{1}, \nonumber \\
&
v^{\uparrow,p_{y}(w)}_{1},v^{\downarrow,p_{y}(w)}_{1},v^{\uparrow,p_{x}(w)}_{2},v^{\downarrow,p_{x}(w)}_{2},v^{\uparrow,p_{y}(w)}_{2},v^{\downarrow,p_{y}(w)}_{2}],
\end{align}
yields the following expression for $\hat{H}$, $\hat{H}_{0}$ and $\delta \hat{H}$
\begin{widetext}
\begin{subequations}
\begin{align}
    \setlength\arraycolsep{7pt}
    \hat{H}=
    \begin{pmatrix}
        \hat{H}_{11}^{(0)}+\delta \hat{H}_{11}& (\hat{H}_{12}^{(0)}+\delta \hat{H}_{12})&0&0&\cdots \\
        (\hat{H}_{12}^{(0)}+\delta \hat{H}_{12})^{T}& \hat{H}_{22}^{(0)}+\delta \hat{H}_{22}&(\hat{H}_{23}^{(0)}+\delta \hat{H}_{23})&\cdots&\cdots\\
        0&(\hat{H}_{23}^{(0)}+\delta \hat{H}_{23})^{T}&\ddots&(\hat{H}_{(\eta-1)\eta}^{(0)}+\delta \hat{H}_{(\eta-1)\eta})&\cdots\\
        0&\vdots&(\hat{H}_{(\eta-1)\eta}^{(0)}+\delta \hat{H}_{(\eta-1)\eta})^{T}&\hat{H}_{\eta\eta}^{(0)}+\delta \hat{H}_{\eta\eta}&\cdots\\
        \vdots&\vdots&\vdots&\vdots&\ddots
    \end{pmatrix}\label{eqn:HGamma}
\end{align}
\begin{align}
    \hat{H}_{\eta\eta}^{(0)}=
    \begin{pmatrix}
        -U_{\eta}&0&0&0&0& \\
        0&(E_{c_{1}}-U_{\eta})&0&0&0& \\
        0&0&(E_{v}-U_{\eta})&0&0& \\
        0&0&0&(E_{v_{1}}-U_{\eta})\bold{\hat{I}}_{\nu}&0&\\
        0&0&0&0&(E_{v_{2}}-U_{\eta})\bold{\hat{I}}_{\nu}&\\
    \end{pmatrix}\label{eqn:H0}
\end{align}
\begin{align} \small
    \delta \hat{H}_{\eta\eta}=
    \begin{pmatrix}
        0&\mathcal{E}_{\eta}d_{c_{1}c}&\mathcal{E}_{\eta}d_{cv}&ib_{54}(\bold{k} \cdot \bold{\Lambda})&i\lambda_{46}(\bold{s}\times \bold{\Lambda})&\\
        \mathcal{E}_{\eta}d_{c_{1}c}&0&0&0&ib^{c_{1}v_{2}}_{16}(\bold{k} \cdot \bold{\Lambda})&\\
        \mathcal{E}_{\eta}d_{cv}&0&0&i\lambda_{15}(\bold{s}\times \bold{\Lambda})&ib_{16}(\bold{k} \cdot \bold{\Lambda})&\\
        -ib_{54}(\bold{k} \cdot \bold{\Lambda})^{T}&0&-i\lambda_{15}(\bold{s}\times \bold{\Lambda})^{\dag}&0&\mathcal{E}_{\eta}d_{v_{1}v_{2}}\bold{\hat{I}}_{\nu}&\\
        -i\lambda_{46}(\bold{s}\times \bold{\Lambda})^{\dag}&-ib^{c_{1}v_{2}}_{16}(\bold{k} \cdot \bold{\Lambda})^{T}&-ib_{16}(\bold{k} \cdot \bold{\Lambda})^{T}&\mathcal{E}_{\eta}d_{v_{1}v_{2}}\bold{\hat{I}}_{\nu}&0& \\
    \end{pmatrix}\label{eqn:deltaH}
\end{align}
\begin{equation}
    \hat{H}_{(\eta-1)\eta}^{(0)}=
    \begin{pmatrix}
        t^{\Gamma}_{cc}&0&0&0&0& \\
        0&0&0&0&0& \\
        0&0&t^{\Gamma}_{vv}&0&0& \\
        0&0&0&0&0& \\
        0&0&0&0&0& \\
    \end{pmatrix}, \delta \hat{H}_{(\eta-1)\eta}=
    \begin{pmatrix}
        0&(t_{cc_{1}}+\delta_{c_{1}c})&(t^{\Gamma}_{cv}+\delta_{cv})&0&0&\\
        (-t_{cc_{1}}+\delta_{c_{1}c})&0&0&0&0& \\
        (-t^{\Gamma}_{cv}+\delta_{cv})&0&0&0&0& \\
        0&0&0&0&(t_{v_{1}v_{2}}+\delta_{v_{1}v_{2}})& \\
        0&0&0&(-t_{v_{1}v_{2}}+\delta_{v_{1}v_{2}})&0& \\
    \end{pmatrix}.\label{eqn:interba}
\end{equation}
\end{subequations}
\end{widetext}
Here, indices $\eta$ and $\kappa$ label layers. The basis of each matrix $\hat{H}_{\eta\kappa}^{(0)}$ and $\delta \hat{H}_{\eta\kappa}$ is the 14-band monolayer InSe basis. In $\hat{H}^{(0)}_{\eta\eta}, $ $U_{\eta}$ is the electrostatic potential in the $\eta^{th}$ layer, $E_{v}$ is the monolayer topmost valence band energy as previously defined in the 2-band model, $E_{v_{1}}$, $E_{v_{2}}$ and $E_{c_{1}}$ are the energies of the $v_{1}$ $v_{2}$ and $c_{1}$ bands, and $\bold{\hat{I}}_{\nu}$ is the identity operator in the space of atomic $p_{x},p_{y}$ orbitals. In $\hat{H}^{(0)}_{(\eta-1)\eta}$, parameters $t^{\Gamma}_{cc}$ and $t^{\Gamma}_{vv}$ are the neighbouring conduction-conduction (valence-valence) interlayer hoppings; no spin index has been included in Eq. (\ref{eqn:interba}) and in Eq. (\ref{eqn:H0}) as all non-zero matrix elements are spin independent. In $\delta \hat{H}_{(\eta-1)\eta}$, $t^{\Gamma}_{cv}$ and $\delta_{cv}$ are the \textit{z}-symmetric and \textit{z}-antisymmetric $c-v$ mixing interlayer hoppings, respectively (see Appendix \ref{app:gammaepsilon}). In $\delta \hat{H}_{\eta\eta}$,  $d_{cv}$, $d_{v_{1}v_{2}}$ and $d_{c_{1}c}$ are the out-of-plane dipole moments (see Fig. \ref{fig:bandsandcharacter}). Coefficients $b_{45}$, $b_{16}$ and $b^{c_{1}v_{2}}_{16}$ are $\bold{k \cdot p}$ mixing terms between $c-v_{1}$, $v-v_{2}$ and $c_{1}-v_{2}$ respectively, while $\lambda_{46}$ and $\lambda_{15}$ are the atomic orbital SOC strengths for $c-v_{2}$ and $v-v_{1}$ spin-flip mixing, with values given in Table \ref{tbl:GWparameters14band}. The latter is included using spin matrices $\bold{s_{x}}$ and $\bold{s_{y}}$. Matrices $\bold{\Lambda_{y}}$ and $\bold{\Lambda_{x}}$ are $1\times2$ matrices $[0,1]$ and $[1,0]$, respectively, operating in the $p_{x},p_{y}$ orbital component of the $v_{1}$ and $v_{2}$ valence bands and $\bold{k} \cdot \bold{\Lambda}\equiv k_{x}\bold{\Lambda_{x}}+k_{y}\bold{\Lambda_{y}}$ and $\bold{s}\times \bold{\Lambda}\equiv \bold{s_{x}}\bold{\Lambda_{y}}-\bold{s_{y}}\bold{\Lambda_{x}}$. \newline
\indent In the absence of interband hoppings, and having neglected the interlayer hoppings between the deeper valence bands $v_{1}$ and $v_{2}$ and between band $c$ and the upper conduction band $c_{1}$, the subband eigenstates formed by $\hat{H}_{0}$ define the orthogonal basis used in the L{\"o}wdin projection. The eigenstates of the $j^{th}$ conduction and valence subband states in this unperturbed Hamiltonian therefore have the form  $\vert c^{j}\rangle=\sum^{\eta=N}_{\eta=1} \alpha^{j}_{\eta} \vert c_{\eta} \rangle$, $\vert v^{j}\rangle= \sum^{\eta=N}_{\eta=1} \beta^{j}_{\eta} \vert v_{\eta} \rangle$, $\vert v^{j}_{1(2)} \rangle=\vert v_{1(2)\eta} \rangle$, $\vert c^{j}_{1} \rangle=\vert c_{1\eta} \rangle$ where $\vert c_{\eta} \rangle$ , $\vert v_{\eta} \rangle$, $\vert v_{1(2)\eta} \rangle$ and $\vert c_{1} \rangle$ are the $c$, $v$, $v_{1(2)}$ and $c_{1}$  monolayer eigenstates in layer $\eta$, respectively. In the following analysis we will only focus on the lowest conduction subband $\alpha^{1}_{\eta}\equiv\alpha_{\eta}$. For the purpose of calculating the SOC coefficient as a function of carrier density, the $v_{1}$, $v_{2}$ and $c_{1}$ subbands are approximated as all being located at $E'_{v_{1}}\equiv E_{v_{1}}-U_{av}$, $E'_{v_{2}}\equiv E_{v_{2}}-U_{av}$ and $E'_{c_{1}}\equiv E_{c_{1}}-U_{av}$ respectively, where $U_{av}$ is the average electrostatic potential per layer. This is due to a small change in the on site electrostatic potential, $\Delta U_{(\eta-1)\eta}=U_{\eta-1}-U_{\eta}$, as compared with the $E_{v_{1}}$, $E_{v_{2}}$ and $E_{c_{1}}$ energy denominators (0.1$-$0.3 eV, as compared to about 3.5 eV for $c$ to $v_{1(2)}$ energy denominator terms and to about 1.4 eV for the $c$ to $c_{1}$ terms). When applying the L{\"o}wdin partitioning method\cite{Lowdinpart,PartLow} (see Appendix \ref{app:Lowdin}), the A block is chosen to act on the $\uparrow$ and $\downarrow$ spin states of the lowest conduction subband and the B block on every other subband in the InSe multiband structure. \newline
\indent In order to obtain the SOC term perturbatively, we account for three effects: an inversion symmetry breaking (such as an electric field or the interlayer pseudopotentials); SOC interband mixing; and $\bold{k \cdot p}$ mixing elements. Consequently, the lowest-order non-zero terms in the perturbation theory have to be third-order in the expansion. Defining,
\begin{align}
    H'_{\rho\omega} \equiv \langle \rho \vert \delta \hat{H} \vert  \omega \rangle,
\end{align}
where $\vert\rho\rangle$ and $\vert\omega\rangle$ are two eigenstates of $\hat{H}_{0}$, the corresponding third-order terms in quasi-degenerate perturbation theory have the form,
\begin{flalign}
\centering
    &
    \Delta H^{(3)}_{mm'}=-\frac{1}{2}\sum_{l,m''}\frac{H'_{ml}H'_{lm''}H'_{m''m'}}{(E_{m'}-E_{l})(E_{m''}-E_{l})}  \nonumber \\
    & -\frac{1}{2}\sum_{l,m'''}\frac{H'_{mm''}H'_{m''l}H'_{lm'}}{(E_{m}-E_{l})(E_{m''}-E_{l})}\nonumber \\
    & 
    +\frac{1}{2}\sum_{l,l'}\frac{H'_{ml}H'_{ll'}H'_{l'm'}}{(E_{m}-E_{l})(E_{m}-E_{l'})}\nonumber \\
    &
    +\frac{1}{2}\sum_{l,l'}\frac{H'_{ml}H'_{ll'}H'_{l'm'}}{(E_{m'}-E_{l})(E_{m'}-E_{l'})},\label{eqn:Lowdin}
\end{flalign}
where the $m,m'$ indices correspond to $\hat{H}_{0}$ subband eigenstates in block A and the $l, l'$ index to any subband eigenstate in block B (see Appendix \ref{app:Lowdin}).
\begin{figure} [t!]
	\includegraphics[width=\columnwidth]{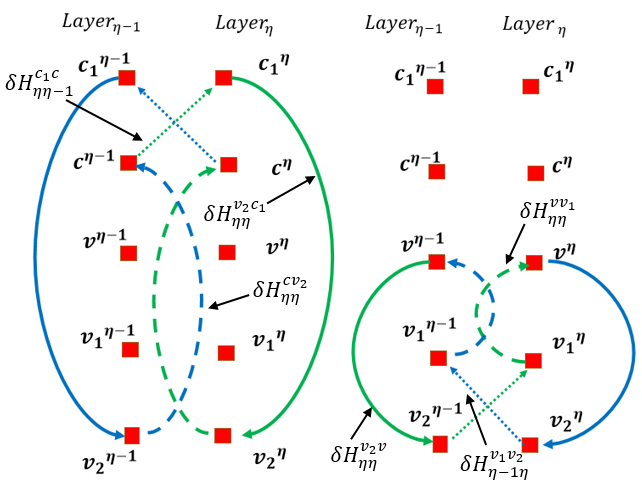}
	\caption{\footnotesize (Left) Feynman diagram of the interlayer spin-flip loops due to the $\gamma$-stacking involving the upper conduction band $c_{1}$. (Right) Feynman diagram of the interlayer spin-flip loops due to the $\gamma$-stacking involving the deeper valence bands $v_{1}$ and $v_{2}$. Such contribution is only relevant for the Dresselhaus SOC in the valence band $v$ as shown in Appendix \ref{app:gammaepsilon}. Doted lines ($\cdots$) label the terms in $\delta \hat{H}$ responsible for inversion symmetry breaking. Dashed lines ($---$) label the intra-atomic SOC mixing between different bands. Solid lines label the $\bold{k\cdot p}$ interband mixing terms in $\delta \hat{H}$. Different colors label pairs of loops that produce competing contributions in the same order of perturbation theory.}\label{fig:gammastackedrashbac1}
\end{figure}
\begin{figure} [t!]
	\includegraphics[width=\columnwidth]{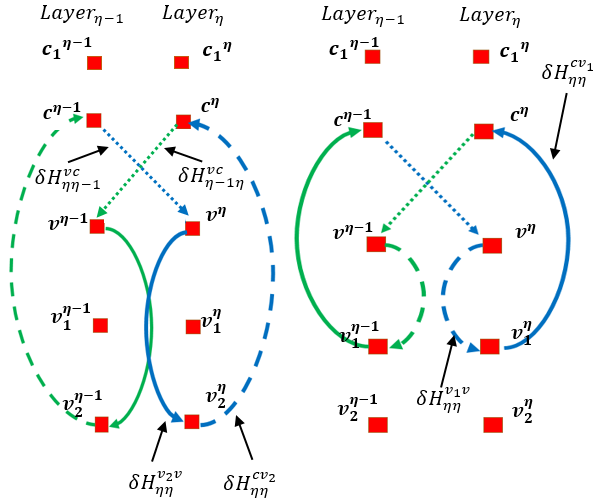}
	\caption{\footnotesize Feynman diagram of the interlayer spin-flip loops due to the $\gamma$-stacking responsible for breaking the $z \rightarrow -z$ symmetry in the $c$ to $v$ hopping parameters $t_{cv}$ and $t_{vc}$. Doted, dashed and solid lines follow the same convention as in Fig. \ref{fig:gammastackedrashbac1}.}\label{fig:gammastackedrashba}
\end{figure}
Energies $E_{m(l)}$ correspond to the energy of the $m^{th}$ or $l^{th}$ eigenstate. Contributions to SOC originate from the 3-step loop Feynman diagrams in Fig. \ref{fig:gammastackedrashbac1}-\ref{fig:dipole}, with spin reversed initial and final states $c^{\uparrow(\downarrow)}$ and $c^{\downarrow(\uparrow)}$. \newline
\indent The Feynman diagrams, originating from the inversion asymmetric parameter $\delta_{cv}$ and $\delta_{c_{1}c}$ in combination with the mixing with deeper valence bands and SOC as shown in Fig. \ref{fig:gammastackedrashbac1} and Fig. \ref{fig:gammastackedrashba}, give a term,
\begin{align} \tiny
    &
   \Delta H'''_{11}=2\Bigg[\sum^{j=N}_{j=1}\sum^{\kappa=N}_{\kappa=1}\Bigg( \frac{b_{54}\lambda_{15}\delta_{cv}}{\Delta E_{cv_{1}}\Delta E_{cv^{j}}}+\frac{b_{16}\lambda_{46}\delta_{cv}}{\Delta E_{cv_{2}}\Delta E_{cv^{j}}}\Bigg) \nonumber \\
   &
   \times\alpha_{\kappa}(\beta^{j}_{\kappa+1}+\beta^{j}_{\kappa-1})\sum^{\xi=N}_{\xi=1}\alpha_{\xi}\beta^{j}_{\xi}+\sum^{\eta=N}_{\eta=1}\Bigg(\frac{b^{c_{1}v_{2}}_{16}\lambda_{46}\delta_{c_{1}c}}{\Delta E_{cc_{1}}\Delta E_{cv_{1}}}\Bigg)\times \\ \nonumber
   &
   \alpha_{\eta}(\alpha_{\eta+1}+\alpha_{\eta-1})\Bigg)\Bigg](\mathbf{\bold{s}\times k}), \label{eqn:delta}
\end{align}
where $\delta_{cv}$ and $\delta_{c_{1}c}$ is the $z$-asymmetric parameters between $c$ and $v$ and between $c_{1}$ and $c$ defined in Eq. (\ref{eqn:interba}) and further discussed in Appendix \ref{app:gammaepsilon}. In the presence of an external electrostatic potential, the signs of $\delta_{cv}$ and $\delta_{c_{1}c}$ become important, as it can be related to placing a single electrostatic gate on one of the surfaces and the orientation (up/down) of externally controlled electric field, $\mathcal{E}_{z}$. \newline
\indent The two diagrams in Fig. \ref{fig:electrostaticU}, give a SOC term in the form of,
\begin{align}
    &
   \Delta H'_{11}=2\Bigg[\sum^{j=N}_{j=1}\sum^{\kappa=N}_{\kappa=1}\Big( \frac{b_{54}\lambda_{15}t^{\Gamma}_{cv}}{\Delta E_{cv_{1}}\Delta E_{cv^{j}}}+\frac{b_{16}\lambda_{46}t^{\Gamma}_{cv}}{\Delta E_{cv_{2}}\Delta E_{cv^{j}}}\Bigg) \nonumber \\
   &
   \times\alpha_{\kappa}(\beta^{j}_{\kappa+1}-\beta^{j}_{\kappa-1})\Bigg(\sum^{\xi=N}_{\xi=1}\alpha_{\xi}\beta^{j}_{\xi}\Bigg)\Bigg](\mathbf{\bold{s}\times k}),
\end{align}
where $\alpha_{\kappa}$ and $\beta_{\kappa}^{j}$ are the components of the lowest conduction subband and the $j^{th}$ valence subband respectively; $\kappa$ labels the layer index. 
\begin{figure} [t!]
	\includegraphics[width=\columnwidth]{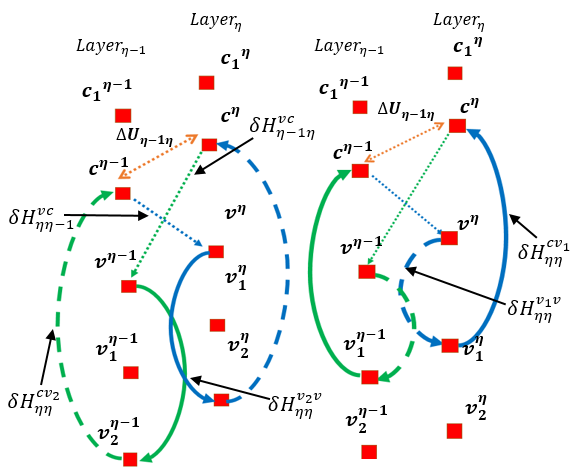}
	\caption{\footnotesize Feynman diagram of the SOC originated from the asymmetry induced by the electrostatic potential distribution $U_{i}$ combined with the $z \rightarrow -z$ symmetric interband hopping parameter $t_{cv}$. Doted, dashed and solid lines follow the same convention as in Fig. \ref{fig:gammastackedrashbac1}.}\label{fig:electrostaticU}
\end{figure}
$\Delta E_{cv^{j}} \equiv E_{c}-E_{v^{j}}$ is the energy difference between the lowest conduction subband and the $j^{th}$ valence subband and $\Delta E_{cv_{1(2)}}\equiv E_{c}-E'_{v_{1(2)}}$ is the energy gap between the lowest conduction subband and the $v_{1}$ and $v_{2}$ subbands located at $E_{v_{1(2)}}-U_{av}$. The loops shown in Fig. \ref{fig:dipole} for the dipolar mixing terms give a SOC term in the form of
\begin{align}
    &
    \Delta H''_{11}= 2\Bigg[\sum^{j=N}_{j=1}\sum^{\kappa=N}_{\kappa=1}\left(\frac{d_{cv}\lambda_{15}b_{54}}{\Delta E_{cv^{j}}\Delta E_{cv_{1}}}+\frac{d_{cv}\lambda_{46}b_{16}}{\Delta E_{cv^{j}} \Delta E_{cv_{2}}} \right) \nonumber \\
    &
    \times(\mathcal{E}_{\kappa}\alpha_{\kappa}\beta^{j}_{\kappa})\Bigg(\sum^{\xi=N}_{\xi=1}\alpha_{\xi}\beta^{j}_{\xi}\Bigg)+\sum^{\eta=N}_{\eta=1}\alpha^{2}_{\eta}\mathcal{E}_{\eta}\times\bigg(\frac{d_{v_{1}v_{2}}b_{54}\lambda_{46}}{\Delta E_{cv_{1}} \Delta E_{cv_{2}}} \nonumber \\
    &
    +\frac{d_{c_{1}c}b^{c_{1}v_{2}}_{16}\lambda_{46}}{\Delta E_{cc_{1}} \Delta E_{cv_{2}}}\bigg)  \Bigg] (\mathbf{\bold{s}\times k}),
\end{align}
where $d_{cv}$ is the matrix element of the out-of-plane dipole operator between the monolayer conduction and valence bands, $d_{v_{1}v_{2}}$ is the out-of-plane dipole moment between $v_{1}$ and $v_{2}$ and $d_{c_{1}c}$ the out-of-plane dipole between $c_{1}$ and $c$.
$\mathcal{E}_{\eta}$ is defined as the electric field in layer $\eta$ and $\Delta E_{cc_{1}}\equiv E_{c}-E'_{c_{1}}$ is the energy difference between the lowest conduction subband and the set of $c_{1}$ subbands located at $E_{c_{1}}-U_{av}$. In accounting for the dipolar terms, some care must be taken in choosing its sign in the few-layer case, as is further explained in Appendix \ref{app:choiceofsign}. \newline
\begin{figure} [t!]
	\includegraphics[width=1.75in]{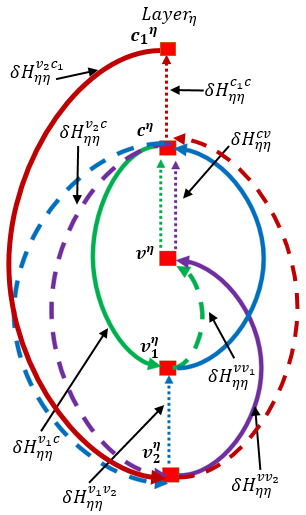}
	\caption{\footnotesize Feynman diagram of the SOC from the dipolar mixing terms. Dots, dashed and solid lines follow the same convention as in Fig. \ref{fig:gammastackedrashbac1}. Different colors label the different 3-step loops included in Eq. (\ref{eqn:Lowdin}) in the same order of perturbation theory.}\label{fig:dipole}
\end{figure}
\indent Combining all these contributions enables us to describe the dependence of SOC strength, $\alpha$, on the number of layers, electric field, and doping in the film as shown in Figs. \ref{fig:carrierdisp},\ref{fig:Rashbacarrier} and \ref{fig:mainresult}. For example, as illustrated in Fig. \ref{fig:Rashbacarrier}, in multilayer InSe in a single-gated FET, doping the device to carrier densities $>10^{13}$cm\textsuperscript{-2} can lead to the compensation of the intrinsic SOC by the contribution of the gate-induced electric field.
\\

\subsection{SOC analysis in I\MakeLowercase{n}S\MakeLowercase{e} films using a quantum well model}

\begin{figure}[t!]
    \centering
    \includegraphics[width=\columnwidth]{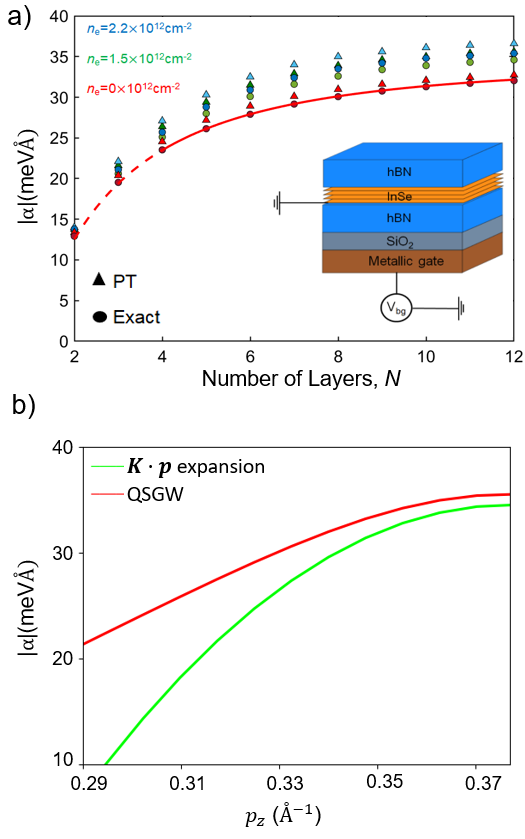}
	\caption{\footnotesize (a) SOC coefficient $\alpha$ computed for InSe films with various thicknesses and carrier densities in a single-gated FET geometry (inset), calculated assuming $\epsilon_{z}$=9.9 for the InSe. The data shown in circles ($\bigcirc$) were obtained by exact diagonalization of the 14-band Hamiltonian in Ref. \onlinecite{ExcitonsGW} and compared with the perturbation theory results obtained by L{\"o}wdin partitioning ($\bigtriangleup$) (note that for N=1, $\alpha=0$ for $n_{e}=0\times 10^{12}$cm\textsuperscript{-2}). Inset shows the usual configuration of a single-gated FET device. The solid and dashed lines indicate the fitted dependence of the Dresselhaus term as a function of the number of layers when the quantum well approximation holds ($N\geq4$) and when it does not respectively. (b) In red, the QSGW-calculated SOC strength as a function of $p_{z}$ in bulk $\gamma$-InSe for the conduction band\cite{Questaal,Questaal_paper}. As $p_{z}$ approaches the band edge located at $p_{z}=\frac{\pi}{a_{z}}=0.378$~\AA$^{-1}$, the SOC strength increases following a quadratic dependence on momentum $p_{z}$ shown in green}$\Bigg( \alpha(p_{z})=\alpha_{\infty}\Big(1-\frac{\chi a_{z}^{2}p_{z}^{2}}{\pi^{2}}\Big)$, where $\alpha_{\infty}=34.5$meV\AA~and $\chi=14.9\Bigg)$. This increasing trend indicates that for greater confinement under a decreasing number of layers, a weaker linear Dresselhaus SOC is expected.
	\label{fig:mainresult} 
\end{figure}

To describe thicker films, it is more practical to use a quantum well model for InSe films \cite{tunelJohanna,subbands2018}. For this, we describe the dispersion of electrons in the $\bold{k}\cdot\bold{p}$ theory expansion near the A-point conduction band edge of bulk InSe as
\begin{align}
    &
   E_{c}(\bold{p},p_{z})=\Bigg(\frac{\hbar^{2}}{2m_{A}}+ \xi p_{z}^{2}a_{z}^{2}\Bigg)p^{2}+\frac{\hbar^{2}p_{z}^{2}}{2m_{Az}} \nonumber \\
    &
    +\alpha_{\infty}\Bigg(1-\frac{\chi a_{z}^{2}p_{z}^{2}}{\pi^{2}}\Bigg)(\bold{s}\times \bold{p}),\label{eqn:bulkSOC}
\end{align}
where $m_{A}$ and  $m_{Az}$ are the in-plane and out-of-plane effective mass at the A-point and the parameters $\xi$ and $\chi$ take into account the anisotropic non-parabolicity of the electron's dispersion characteristic for layered systems. In Fig. \ref{fig:mainresult}(b) we show the $p_{z}$-dependence (around the A-point) of the linear in $k_{x},k_{y}$ spin-orbit coupling computed by DFT for bulk InSe using QSGW approach\cite{ExcitonsGW,Questaal_paper}, to compare with the SOC form in Eq. (\ref{eqn:bulkSOC}). This has to be complemented with the generalised Dirichlet-Neumann boundary conditions for the quantum well wavefunction $\Psi(z)$ at the encapsulating interfaces,
\begin{align}
    \Psi\pm\nu a_{z}\partial_{z}\Psi=0, \quad \nu \approx 1.42.
\end{align}
The latter determines the values for the wave numbers of the electron's standing waves,
\begin{align}
   p_{z}= \frac{n\pi}{(N+2\nu)a_{z}},
\end{align}
which determines the subband and layer-number-dependence of the subband mass and SOC parameter,
\begin{align} 
    &
    \frac{1}{m_{n}}\approx\frac{1}{m_{A}}\Bigg(1-\frac{6.2n^{2}}{(N+2\nu)^{2}}\Bigg),\\ \nonumber
    &
    \alpha_{n|N}(n<<N)\approx\alpha_{\infty}\Bigg(1-\frac{\chi}{(N+2\nu)^{2}}\Bigg). \label{eqn:alphaN}
\end{align}
By fitting $\alpha_{1|N}$ described in Eq. (\ref{eqn:formSOCexp}) to the values of the lowest subbands SOC strength in Fig. \ref{fig:mainresult} we find that $\alpha_{\infty}$=34.5 meV{\AA} and $\chi$=14.9 respectively. Additionally, the results of the calculations, performed in the same films subjected to an electric field $\mathcal{E}_{z}$ perpendicular to the layers shown in Fig. \ref{fig:RashbaEfield} and \ref{fig:carrierdisp} show an approximately linear SOC strength dependence on $\mathcal{E}_{z}$. We describe the latter as
\begin{align}
    \alpha(N,\mathcal{E}_{z})=\alpha(N)-\mathcal{E}_{z}\aleph(N),
\end{align}
with the vales of $\aleph(N)$ for $N\geq 2$ shown in the inset of Fig. \ref{fig:RashbaEfield}. Further to the DFT calculations for the few-layer case, in Fig. \ref{fig:plainGW}(b) we use previous quasiparticle self-consistent GW (QSGW) calculations for bulk $\gamma$-InSe\cite{ExcitonsGW,Questaal_paper} to extract the $k_{z}$-dependence of the coefficient of the linear component of SOC for small in-plane momentum near $k_x=k_y=0$, for both the conduction and valence bands. This shows that as $k_z$ approaches the bulk band edge (located at $k_z=\pi/a_z$) the SOC strength increases, implying that as $k_z$ is restricted by confinement in thin films of InSe, the SOC strength can be expected to decrease from its bulk value, with smaller strengths for thinner films. 

\section{Magnetotransport studies of I\MakeLowercase{n}S\MakeLowercase{e} films in the FET geometry and their comparison with theory}\label{section:magnetoconductivity}
\begin{figure}[b!]
    \centering
    \includegraphics[width=\columnwidth]{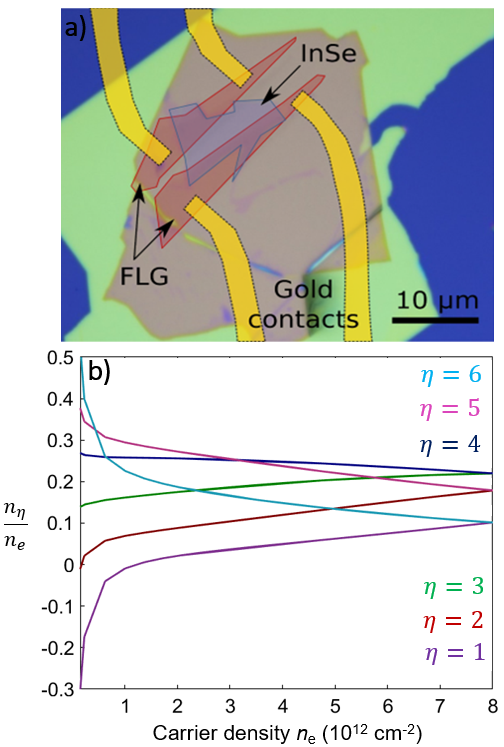}
	\caption{\footnotesize (a) Optical photograph of an encapsulated InSe flake (light blue) equipped
    with few-layer graphene (FLG) contacts (red). Yellow polygons illustrate gold
    leads contacting FLG. Green colour corresponds to the bottom hBN flake deposited on top of an oxidized Si wafer (dark blue). (b) Calculated charge density distribution along the different layers in the dual-gated device under study. At a carrier concentration of $n_{e}=8\times10^{12}$ cm\textsuperscript{-2}, the distribution of charges becomes $z \rightarrow-z$ symmetric as the top plate carrier density is fixed at $n_{tg}=4\times10^{12}$ cm\textsuperscript{-2}.}\label{fig:device&charges}
\end{figure}
\begin{figure} [t!]
    \centering
    \includegraphics[width=\columnwidth, height=6.5in]{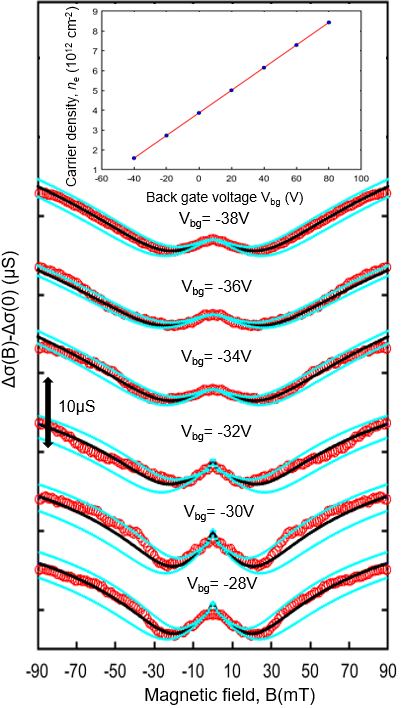}
	\caption{\footnotesize Weak antilocalization feature in conductivity measured in a 6-layer dual-gated InSe device with corresponding optimal fits (Black). Carrier densities were measured in the range from $1.7-2.2 \times 10^{12}$ cm\textsuperscript{-2} in steps of $0.1 \times 10^{12}$ cm\textsuperscript{-2}. Blue indicates the upper and lower bound fits of the corrections to magnetoconductivity. Top inset shows the carrier density at each back gate voltage obtained from Hall-effect measurements. The finite carrier density at $V_{bg}$=0 is due to the applied top gate voltage corresponding to $V_{tg}=8$ V. The linear relation between the carrier density and the back gate voltage for a fixed top gate of $V_{tg}=8$ V was found to be $n_{e}=\Upsilon(V_{bg}-V'_{bg(V_{tg}=\textrm{8V})})$ where $V'_{bg(V_{tg}=\textrm{8V})}=-67.6$V and $\Upsilon=5.71\times10^{10}$ V\textsuperscript{-1}cm\textsuperscript{-2}. }\label{fig:walmeasurements}
\end{figure}
\begin{figure}[t!]
    \includegraphics[width=\columnwidth]{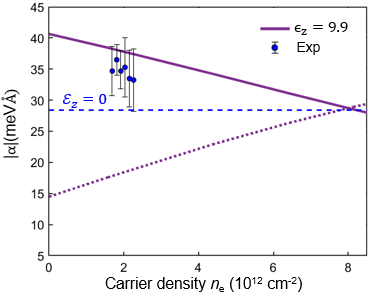}
	\caption{\footnotesize SOC coefficient $\alpha$ as experimentally extracted from weak antilocalization measurements of the dual-gated 6 layer device\cite{bandurin2017high}, compared to the value obtained in the self-consistent calculation. The blue dashed line indicates the value of $\alpha$ in the absence of any electrostatic gating and doping. The same notation for the solid and dotted lines is used as in Fig. \ref{fig:RashbaEfield},\ref{fig:Rashbacarrier} and \ref{fig:carrierdisp}.}\label{fig:comparison}
\end{figure}

In order to probe the nature of SOC in InSe, we fabricated a dual-gated multiterminal 6 layer $\gamma$-InSe device using mechanical exfoliation and hexagonal boron nitride (hBN) encapsulation, which were carried out in an inert atmosphere of
a glovebox\cite{Quantumhetero}. Such encapsulation was needed to protect air-sensitive InSe flakes from the environment (see Fig. \ref{fig:device&charges}(a)). In addition, electrical contact to InSe was provided by few-layer graphene (FLG) flakes which in turn were connected to metal leads by standard nanofabrication techniques as illustrated in Fig. \ref{fig:device&charges}(a) (see Ref. \onlinecite{bandurin2017high} for further details). The gate-tunable work function of graphene ensured ohmic contacts between FLG and 2D InSe\cite{multi} and thus enabled us to explore InSe properties using conventional four-terminal measurements. To characterize the fabricated device, we first measured its longitudinal resistivity, $\rho_{xx}$, as a function of gate-induced carrier density, $n_{e}$. The latter was obtained via Hall-effect measurements that provided full $n_{e}(V_{bg})$ dependence presented in Fig. \ref{fig:walmeasurements}. In contrast to earlier studies of the quantum Hall effect in InSe/graphene interfaces\cite{GiantQHE}, the perfectly linear $n_{e}$ vs $V_{bg}$ trend shown in the inset of Fig. \ref{fig:walmeasurements} does not indicate any substantial charge transfer from the InSe to the gating surface. Using Drude formula we determined the mean free path of charge carriers, $\lambda$, and respective scattering time, $\tau$, important parameters critical for
further analysis. The effective mass for the lowest conduction subband used to extract $\tau$ was $m_{c}=0.12m_{e}$, obtained from an accurate calculation of the bulk effective mass accounting both for electron-electron and electron-phonon interaction effects in the bulk conduction band.\cite{Renor}. 
\begin{figure}[t!]
    \includegraphics[width=\columnwidth]{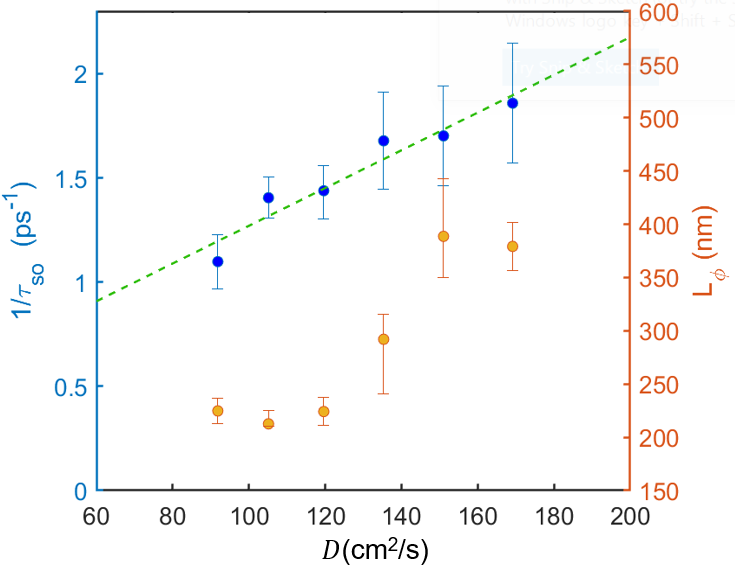}
	\caption{\footnotesize Inverse spin relaxation time and phase relaxation length \emph{vs} diffusion coefficient. The proportionality relation between the diffusion coefficient obtained by varying the carrier density $n_{e}$ and the inverse of the spin relaxation time indicates Dyakonov-Perel mechanism of spin relaxation.}\label{fig:DPfigure}
\end{figure}
\newline
\indent An experimental manifestation of the SOC strength can be found in the weak antilocalization (WAL) corrections to magnetoconductance\cite{Rashba_theory_exp,Pikus,HLNpaper,AllanWL} produced by the interference of electron waves propagating along closed loops of random walks\cite{AltschulerWL,AltschulerAronov}. Such behavior has been observed in recent studies of few-layer single-gated GaSe\cite{GaSe_wal} and InSe\cite{otherInSe_wal,Zeng2018}. \newline
\indent In Ref. \onlinecite{Zeng2018}, the fitting procedure used to extract the SOC strength from the corrections to magnetoconductance was the formalism developed by Hikami, Larkin and Nagaoka \cite{HLNpaper} for systems where the spin relaxation mechanism is dominated by scattering with magnetic impurities \cite{Elliot,Yafet}. As the $\gamma$-stacked phase in InSe is non-centrosymmetric and therefore the spin relaxation mechanism is expected to be Dyakonov-Perel, their extracted spin relaxation parameters from WAL fits were overestimated.
In Ref. \onlinecite{otherInSe_wal} the enhancement of the SOC as compared to our estimated bulk SOC strength value at the band edge ($\alpha_{\infty}\approx$ 34 meV{\AA}) is a result of an impurity deposition layer formed at the interface of the suspended device; this forms a sharp potential barrier at the interface and therefore increases the SOC strength. \newline
\indent From our weak antilocalization measurements, the spin and phase relaxation times can be obtained by fitting the corrections to conductivity with respect to these two parameters in the range of magnetic fields where the minimum in magnetoconductance appears. \newline
\indent The WAL corrections to the conductivity of the 6-layer device with the same characteristics as reported in Ref. \onlinecite{bandurin2017high} were measured as a function of the magnetic field with 1~mT magnetic field step. As shown in Fig. \ref{fig:walmeasurements}, at magnetic fields 10-30~mT, a clear minimum in the magnetoconductance is observed.
The corrections to conductivity $\Delta \sigma(B)-\Delta \sigma(0)$ were measured in the range of 0$-$90~mT, and both the spin and phase relaxation time were fitted with the formalism developed by Iordanskii, Larkin and Pitaevskii\cite{ILPantilocalization,KnapILP} (ILP) for systems where the lack of inversion symmetry leads to the electron's spin precessing and to relaxation by Dyakonov-Perel mechanism. Such formalism was used for carrier densities $<2\times 10^{12}$cm\textsuperscript{-2}; above that carrier densities, the assumption of the ILP formalism that the precession angle $\phi=\Omega\tau<<1$ ($\Omega$ being the spin precession frequency and $\tau$ the momentum relaxation time), and that the magnetic field $B<<B_{tr}$ (where $B_{tr}\equiv \frac{\hbar}{2e\lambda^{2}}$ and $\lambda$ is the mean free path) breaks down. 
The spin precession frequency $\Omega$ is then related to the spin-orbit coupling strength $\alpha$ through the simple relation $\Omega=\alpha k_{F}$ where $k_{F}$ is the Fermi momentum. In these cases, we employ the approach developed by Golub\cite{highmobilityWAL,Golub-Glazov}, which goes beyond the diffusion approximation for arbitrarily large precession angles and for magnetic fields comparable to the transport field $B_{tr}$.\newline
\indent For the magnetoconductance fits performed at carrier densities $n_{e}\geq 2\times10^{12}$ cm\textsuperscript{-2}, the non-backscattering corrections to conductivity were found to be negligible, and therefore corrections to conductivity only came from the backscattering loops,
\begin{widetext}
\begin{equation} \small
    \sigma_{back}=-\frac{e^{2}}{2 \pi^{2} \hbar} \left(\frac{\lambda}{l_{B}} \right)^{2}\sum^{\infty}_{N'=0} \Big(Tr\left[\hat{A}^{3}_{N'}(\hat{I}-\hat{A}_{N'})^{-1}\right]-\frac{P^{3}_{N'}}{1-P_{N'}}\Big),\label{eqn:ltil}
\end{equation}
\begin{equation} \small
    \hat{A}_{N'}\equiv \begin{pmatrix}
                P_{N'-2}-S^{(0)}_{N'-2} &R^{(1)}_{N'-2}&S^{(2)}_{N'-2} \\
                R^{(1)}_{N'-2}&P_{N'-1}-2S^{(0)}_{N'-1}&R^{(1)}_{N'-1} \\
                S^{(2)}_{N'-2}&R^{(1)}_{N'-1}&P_{N'}-S^{(0)}_{N'} \\
        \end{pmatrix}, \nonumber
\end{equation}
\begin{subequations}
\begin{equation} \small
    P_{N'} \equiv \frac{l_{B}}{\lambda}\int^{\infty}_{0} \exp\Bigg(- \frac{l_{B}}{\tilde{l}}x-\frac{x^{2}}{2} \Bigg) L_{N'}(x^{2}) dx, \nonumber
\end{equation}
\begin{equation} \small
    S^{(\mu)}_{N'}\equiv \frac{l_{B}}{\lambda}\sqrt{\frac{N'!}{(N'+\mu)!}}\int^{\infty}_{0} \exp\Bigg(- \frac{l_{B}}{\lambda}x-\frac{x^{2}}{2}\Bigg)x^{\mu}L^{\mu}_{N'}(x^{2})
    \sin^{2} \left(\Omega \tau \frac{l_{B}}{\lambda} \right) dx, \nonumber
\end{equation}
\begin{equation} \small
    R^{(\mu)}_{N'}\equiv \frac{l_{B}}{\sqrt{2}\lambda}\sqrt{\frac{N'!}{(N'+\mu)!}}\int^{\infty}_{0} \exp\Bigg(- \frac{l_{B}}{\lambda}x-\frac{x^{2}}{2}\Bigg) x^{\mu}L^{\mu}_{N'}(x^{2})
    \sin \left( 2\Omega \tau \frac{l_{B}}{\lambda}\right) dx, \nonumber
\end{equation}
\end{subequations}
\end{widetext}
Here, $l_{B}\equiv\sqrt{\frac{\hbar}{eB}}$ is the magnetic length, and in Eq. (\ref{eqn:ltil}), $\tilde{l}$ is defined as $\tilde{l}\equiv \frac{\lambda}{1+\frac{\tau}{\tau_{\phi}}}$ where $\tau_{\phi}$ is the phase relaxation time. The precession frequency is related to the spin relaxation time $\tau_{SO}$ through $\frac{1}{\tau_{SO}}=2\Omega^{2}\tau$. As done previously with the ILP formalism, both the phase and spin relaxation times were taken as fitting parameters. In Fig. \ref{fig:DPfigure}, the inverse proportionality between the spin relaxation time and the diffusion coefficient $D$ confirms that the spin relaxation mechanism is Dyakonov-Perel\cite{dyakonov1972spin,dyakonovPerel}. From $\tau_{SO}$, the SOC coefficient is extracted and compared with our theoretical calculation in Fig. \ref{fig:comparison}. In Fig. \ref{fig:comparison} the SOC coefficient at different carrier densities was calculated at the experimentally established dielectric constant $\epsilon_{z}=9.9$ for InSe\cite{dielectric}. Very good agreement was found between the calculated SOC coefficient and the experimentally extracted SOC strength. Furthermore, by looking at the two different branches originated from the orientation of the crystal being parallel or antiparallel to the applied electric field, it was found that at a carrier density of $n_{e}=8\times 10^{12}$ cm\textsuperscript{-2} the two branches converged at a single point. This indicates no dependence neither on crystal orientation nor on electrostatic profile. As shown in Fig. \ref{fig:device&charges}, at that exact carrier density, the electrostatic profile is expected to be $z\rightarrow-z$ symmetric and therefore the only contribution to the SOC must originate from the intrinsic $z \rightarrow -z$ asymmetry of the crystal (see comparison in Fig. \ref{fig:comparison} with SOC strength at zero electric field).
\begin{widetext}
\begin{figure*}[t!]
    \centering
    \includegraphics[width=7.3in, height=3.7in]{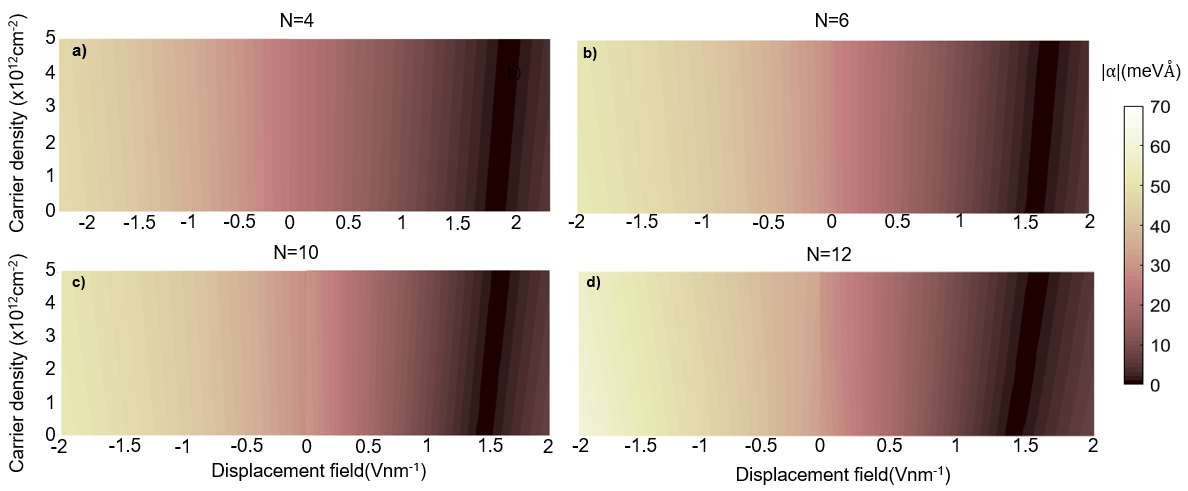}
	\caption{\footnotesize (a-d) SOC strength as a function of the displacement field and carrier density for different number of layers. Crystal orientation is chosen such that the applied displacement field counteracts the Dresselhaus SOC when the displacement field is positive. The dark black lines indicate the disappearance of SOC due to the application of a displacement field which compensates the SOC from the intrinsic lack of inversion symmetry in the different multilayers.}\label{fig:color}
\end{figure*}
\end{widetext}

\section{Conclusion}\label{section:conclusion}

\indent Overall, the description of SOC strength (as a function of the number of layers and the applied electric field piercing the multilayer film) obtained using the few-layer HkpTB study and a quantum well model give the matching results, and the theoretically computed SOC strengths are compared with the results of weak antilocalization measurements on dual-gated multilayer InSe films showing a good agreement between theory and experiment.\\
\indent The size of SOC constant we compute for InSe films with 2-10 layers thickness is comparable to the SOC strength in quantum wells of conventional semiconductors, such as GaAs, InAs, HgTe. What makes 2D InSe different from those spintronic systems is that the SOC strength in it can be tuned over a wide range. Additionally contribution originating from the asymmetry of an hBN/InSe interface was analysed
and shown to be negligible (as compared with the intrinsic SOC in the film) for InSe encapsulated in hBN both on top and in the bottom, and also to decay as $N^{-3}$. Moreover we demonstrate that spin-orbit coupling strength for electrons near the conduction band edge in few-layer $\gamma$-InSe films can be tuned over a wide range, from $\alpha$=0 to $\alpha\approx70\text{meV\AA}$. This tunability illustrated in Fig. \ref{fig:color} for the films of various thicknesses is the result of a competition between film-thickness-dependent intrinsic and electric-field-induced SOC, potentially, allowing for electrically switchable spintronic devices. As shown in Fig. \ref{fig:color} and Fig. \ref{fig:RashbaEfield}, displacement fields in the range of 1-2 Vnm\textsuperscript{-1} can turn the SOC on and off.
\begin{acknowledgements}
The authors thank  M. Potemski, Y. Ye, J. Lischner, A. Mirlin, V. Enaldiev, K.W. Song, S. Slizovskiy, and N. D. Drummond for discussions. We also thank S.J. Liang and F. Miao for sharing the data in Ref. \onlinecite{Zeng2018}. This work made use of the CSF cluster of the University of Manchester and the N8 Polaris service, the use of the ARCHER national UK supercomputer (RAP Project e547), and the Tianhe-2 Supercomputer at NUDT. We acknowledge support from  EPSRC CDT Graphene NOWNANOEP/L01548X, ERC Synergy Grant Hetero2D, grant EP/N010345, Lloyd Register Foundation Nanotechnology grant, European Quantum Technology Flagship Project 2D-SIPC, and Core 3 European Graphene Flagship Project. Research data are available from the authors on request.\\

\end{acknowledgements}

\appendix

\section{Determination of parameters $\delta_{cv}$, $\delta_{c_{1}c}$ and $\delta_{v_{1}v_{2}}$ from bulk SOC}\label{app:gammaepsilon}

\begin{figure}[h!]
    \includegraphics[width=\columnwidth]{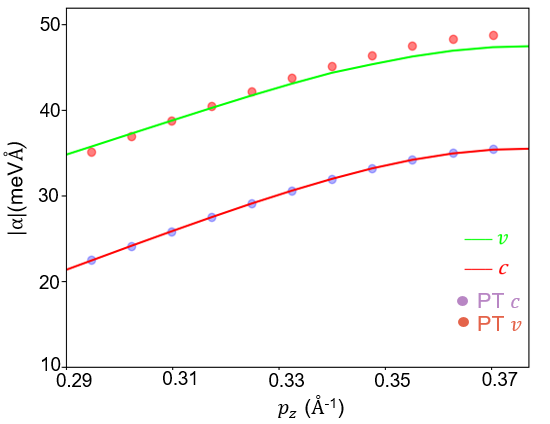}
	\caption{\footnotesize (Solid) Bulk SOC as a function of $p_{z}$ for $c$ and $v$ bands. (Dots) SOC strength at different $p_{z}$ obtained from the perturbative analysis in Eq. (\ref{eqn:Blochc}) and (\ref{eqn:Blochv}).}\label{fig:Blochptex}
\end{figure}

For the calculation of subband energies and dispersions, it was sufficient to approximate the interlayer hops as being entirely between the inversion symmetric sublattices of selenium atoms on the outside of each layer. This causes the hops to be inversion symmetric, which when combined with the opposite \textit{z}-symmetries of the monolayer conduction and valence under $\sigma_h$ reflection (i.e. $z\rightarrow-z$ symmetry) gives $t_{cv}=-t_{vc}$, $t_{c_{1}c}=-t_{cc_{1}}$ and $t_{v_{2}v_{1}}=-t_{v_{1}v_{2}}$. It is transparent from Eq. (\ref{eqn:formSOC}) that inversion symmetry would prohibit the existence of extrinsic SOC. Consequently, we require terms in our model which break inversion symmetry (such as an applied electric field or the interlayer pseudopotentials arising from the $\gamma$-stacking\cite{zhou2019spin}). The indium atoms provide such an asymmetry-in the $\gamma$ stacking there is a vertically opposite interlayer In/Se pair heading in one direction along the \textit{z}-direction, while in the other direction the indium atom is opposite an empty space in the adjacent layer. In the $\bold{k} \cdot \bold{p}$ model, the effect of this symmetry breaking is to give $t_{cv}$, $t_{c_{1}c}$ and $t_{v_{1}v_{2}}$ slightly different magnitudes as compared with $-t_{cv}$, $-t_{c_{1}c}$ and $-t_{v_{1}v_{2}}$, so we define three new parameters: $2\delta_{cv}\equiv t_{cv}$+$t_{vc}$ $2\delta_{c_{1}c}\equiv t_{c_{1}c}$+$t_{cc_{1}}$ and $2\delta_{v_{1}v_{2}}\equiv t_{v_{1}v_{2}}$+$t_{v_{2}v_{1}}$. In order to obtain the parameters $\delta_{cv}$ and $\delta_{c_{1}c}$ relevant for the analysis of the Dresselhaus SOC in the conduction band, the linear SOC splittings at each individual $k_{z}$ are obtained from the QUESTAAL package by linearly fitting the energy differences between the two spin split bands (see Fig. \ref{fig:Blochptex}). Firstly, the parameters $\delta_{cv}$ and $\delta_{c_{1}c}$ were fitted for the $\alpha$ vs $k_{z}$-dependence of band $c$ (red curve in Fig. \ref{fig:Blochptex}), and then the $\delta_{v_{1}v_{2}}$ parameter was fitted from the $\alpha$ vs $k_{z}$-dependence of band $v$ (green curve in Fig. \ref{fig:Blochptex}). Using the same perturbative analysis as in Section \ref{section:perturbationalaysis} in the bulk limit, the Dresselhaus SOC at each $k_{z}$ is obtained both for the $c$ and $v$ bands respectively, namely
\begin{align}
    &
    \alpha_{c}(p_{z})=4\cos{(p_{z}a_{z})}\Bigg(\frac{\delta_{cv}b_{54}\lambda_{15}}{\Big(E_{c}-E_{v}\Big)\Big(E_{c}-E_{v_{1}}\Big)}+ \nonumber \\
    &
    \frac{\delta_{cv}b_{16}\lambda_{46}}{\Big(E_{c}-E_{v}\Big)\Big(E_{c}-E_{v_{2}}\Big)}+\frac{\delta_{c_{1}c}b^{c_{1}v_{2}}_{16}\lambda_{46}}{\Big(E_{c}-E_{c_{1}}\Big)\Big(E_{c}-E_{v_{2}}\Big)}\Bigg) \label{eqn:Blochc}
\end{align}
and
\begin{align}
    &
    \alpha_{v}(p_{z})=4\cos{(p_{z}a_{z})}\Bigg(\frac{\delta_{cv}b_{54}\lambda_{15}}{\Big(E_{v}-E_{c}\Big)\Big(E_{v}-E_{v_{1}}\Big)}+ \nonumber \\
    &
    \frac{\delta_{cv}b_{16}\lambda_{46}}{\Big(E_{v}-E_{c}\Big)\Big(E_{v}-E_{v_{2}}\Big)}+\frac{\delta_{v_{1}v_{2}}b_{16}\lambda_{15}}{\Big(E_{v}-E_{v_{1}}\Big)\Big(E_{v}-E_{v_{2}}\Big)}\Bigg).\label{eqn:Blochv}
\end{align}
where $p_{z}=\frac{\pi}{a_{z}}-k_{z}$. The fitting parameters considered are the terms $\delta_{cv}$, $\delta_{c_{1}c}$, $\delta_{v_{1}v_{2}}$ and $\lambda_{46}$ as the 14-band fit applied to the InSe bulk dispersion did not account for any them. The optimal parameters found in order to fit the spin splitting vs $k_{z}$-dependence in the vicinity of the band edge where perturbation theory is best applicable were $\delta_{cv}=0.014$ eV, $\delta_{c_{1}c}=0.022$ eV, $\delta_{v_{1}v_{2}}=-0.001$ eV and $\lambda_{46}=-0.09$eV.  
\section{Determination of the signs of $d_{cv}$, $d_{v_{1}v_{2}}$ and $d_{c1c}$} \label{app:choiceofsign}

While on their own the signs of $d_{cv}$ and $t_{cv}$ may be chosen arbitrarily through an appropriate choice of basis in the monolayer Hamiltonian, the product of $d_{cv}$ and $t_{cv}$ does not have such degree of freedom. In order to determine the relative signs of the different dipole moments, it is necessary to look at their $k$-dependence as we move away from the $\Gamma$-point. In considering the conduction to valence band interlayer hopping (both the \textit{z}-symmetric and \textit{z}-antisymmetric) as a perturbation to our conduction or valence subband wavefunctions, the $k$-dependence of the bilayer valence band dipole moment follows easily as:
\begin{equation}
\langle v_{2L}\vert ez \vert v_{2L} \rangle=2\delta_{cv}\left[\frac{d_{cv}}{E_{g2L}}+\frac{t_{cv}ea_{z}}{2E_{g2L}E_{g2L}'}\right]
\end{equation}
for the valence band, and 
\begin{equation}
\langle c_{2L} \vert ez \vert c_{2L} \rangle=-2\delta_{cv}\left[\frac{d_{cv}}{E_{g2L}}+\frac{t_{cv}ea_z}{2E_{g2L}E_{g2L}''}\right]
\end{equation}
for the bilayer conduction band. In the above equation, $d_{cv}=\vert \langle c|ez| v \rangle \vert=1.68~e\mathrm{\AA}$ is the matrix element of the out-of-plane dipole operator between the monolayer conduction and valence bands and $v_{2L}$ and $c_{2L}$ are the topmost valence subband and lowest conduction subband wavefunctions in a bilayer system at the $\Gamma$-point. $E_{g2L}=E_{c}-E_{v}-(t_{cc}-t_{vv})$, $E'_{g2L}=E_{c}-E_{v}+t_{cc}+t_{vv}$, and $E''_{g2L}=E_{c}-E_{v}-(t_{cc}+t_{vv})$ are the energy differences between the bilayer bands in the absence of the interband hoppings.
\begin{figure} [t!]
    \centering
    \includegraphics[width=\columnwidth]{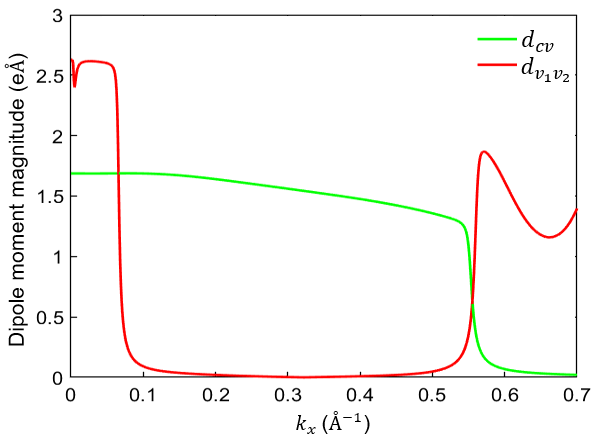}
	\caption{\footnotesize Dipole moments between monolayer bands $c$ and $v$ ($d_{cv}$) and between monolayer bands $v_{1}$ and $v_{2}$ ($d_{v_{1}v_{2}}$) computed using the tight-binding model in Ref. \onlinecite{magorrian2016electronic}.}\label{fig:dipolesk}
\end{figure}
Comparison of these expressions with the signs of the quantities calculated using DFT gives, for a choice of positive $t_{cv}$ and negative $d_{cv}$, a positive $\delta_{cv}$ when the $+z$ direction is chosen such that the vertical In-Se interlayer pair in the interface between two layers the Se atom lies above the In atom in the $\gamma$-stacking. Conversely, a negative $\delta_{cv}$ is obtained for the opposite orientation. On calculating perturbatively the value of $d_{cv}$ at a finite $k$, the following result is obtained
\begin{align} \small
    d_{cv}(k)\equiv\langle v|ez|c \rangle=\langle v_{0} \vert ez \vert c_{0} \rangle+\frac{b_{54}b_{16}k^{2}d_{v_{1}v_{2}}}{\Delta E_{v_{1}c}\Delta E_{v_{2}v}}.
\end{align}
By looking at the negative trend of $|d_{cv}|$ and the hybrid $\bold{k \cdot p}$  tight-binding values quoted in Table \ref{tab:ml_kp_parameters}, it is transparent that if $d_{cv}$ is positive $d_{v_{1}v_{1}}$ is as well positive. Furthermore, if $d_{cv}$ is negative, the value of $d_{v_{1}v_{2}}$ should be negative as well. In order to find the sign of the dipole moment $d_{c_{1}c}$ a similar perturbative analysis is applied for $d_{v_{1}v_{2}}$,
\begin{align}
    &
    d_{v_{1}v_{2}}(k)\equiv\langle v_{1} | ez | v_{2} \rangle=\langle v_{1,0} \vert ez \vert v_{2,0} \rangle+ \frac{b_{54}b_{16}k^{2}d_{cv}}{\Delta E_{v_{1}c}\Delta E_{v_{2}v}}\\
    &
    +\frac{b_{54}b^{c_{1}v_{2}}_{16}k^{2}d_{c_{1}c}}{\Delta E_{v_{1}c}\Delta E_{v_{2}c_{1}}}. \nonumber
\end{align}
In comparing the red and the green curve in Fig. \ref{fig:dipolesk}, the much more pronounced steepness of the red curve as compared to the green curve at low values of $k$ indicates that $d_{c_{1}c}$ must be negative for a positive $d_{v_{1}v_{2}}$ and vice-versa.

\section{Interfacial contribution to multilayer InSe SOC}

\begin{figure}[t!]
	\includegraphics[width=\columnwidth]{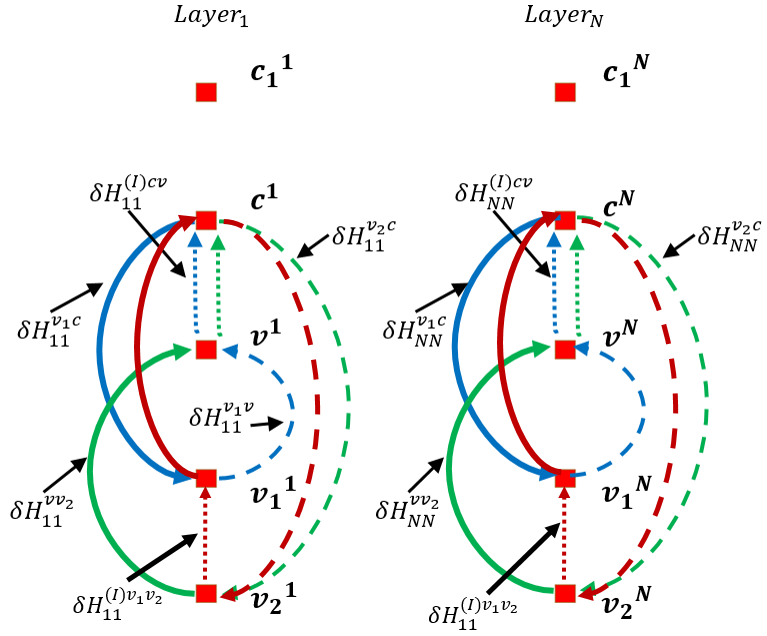}
	\caption{\footnotesize Feynman diagram of the interlayer spin-flip loops due to the interfacial electric fields experienced by the electrons in the outer Se orbitals of the $1^{st}$ and $N^{th}$ layer. Dots, dashed and solid lines follow the same convention as in Fig. \ref{fig:gammastackedrashbac1}.}\label{fig:InterfaceFeynmann}
\end{figure}

\indent In addition to the crystalline and the electrostatically induced $z\rightarrow-z$ asymmetry, few-layer InSe is a material sensitive to interfacial effects due to its limited thickness. Such effects may have an impact in the SOC strength of multilayer InSe and must therefore be taken into consideration\cite{dipoleinterface}. The same two InSe-hBN configurations used for the analysis of interfacial effects in bilayer InSe shown in Table \ref{tab:InSe_hBN_shifts} (configuration 1 and 2) were also used for the calculation of the interface-induced SOC in multilayer InSe as their contribution in the absence of an external electrostatic potential is only dependent on the encapsulating substrates and on the film thickness. Interface effects are taken into account by adding into the multilayer Hamiltonian two additional contributions identical to Eq. (\ref{eqn:H0bilinter}). Firstly, bands $c$ and $v$ with a relevant Se $p_{z}$ orbital composition, experience in the outer layers a shift in energy due to the interaction with the $p_{z}$ orbitals of the encapsulating hBN. Therefore, an additional energy shift is added to the $c$,$v$,$v_{1}$ and $v_{2}$ bands of the $1^{st}$ and the $N^{th}$ layer.
\begin{figure}[b!]
    \centering
    \includegraphics[width=\columnwidth]{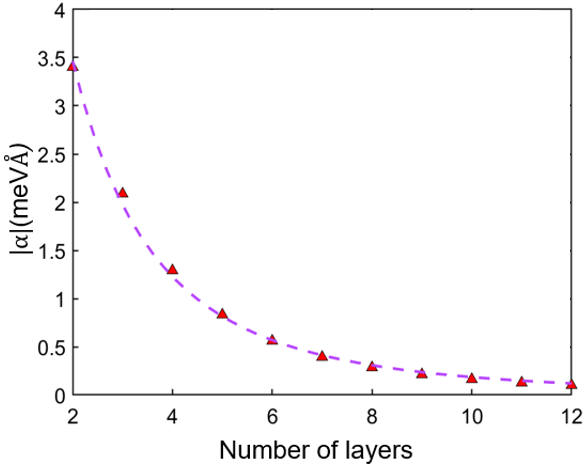}
	\caption{\footnotesize Interfacial SOC as a function of the number of layers in the absence of an externally applied electric field $\mathcal{E}_{z}$. (Dashed) Fit of the interfacial SOC strength as a function of the number of layers. A $\frac{1}{(N+2\nu)^{3}}$ dependence is expected from the quantum well model presented in Ref. \onlinecite{subbands2018}.}
	\label{fig:interfaceQW} 
\end{figure}\\
\indent Additionally, the hBN interfaces break $z\rightarrow-z$ symmetry in the outer layers mixing bands with opposite $z$-parity but identical in-plane symmetries. The following perturbative term accounting for all these  effect is introduced in the multilayer Hamiltonian,
\begin{align}
    &
    \delta\hat{H}^{(I)}_{11(NN)}=\\ \nonumber
    &\begin{pmatrix}
        \Delta E_{c1(N)}&0&\pm\Upsilon^{t/b}_{cv}&0&0&\\
        0&\Delta E_{v1(N)}&0&0&0&\\
        \pm\Upsilon^{t/b}_{cv}&0&0&0&0&\\
        0&0&0&\Delta E_{v_{1}1(N)}\bold{\hat{I}}_{\nu}&\pm\Upsilon^{t/b}_{v_{1}v_{2}}\bold{\hat{I}}_{\nu}&\\
        0&0&0&\pm\Upsilon^{t/b}_{v_{1}v_{2}}\bold{\hat{I}}_{\nu}&\Delta E_{v_{2}1(N)}\bold{\hat{I}}_{\nu}& \\\label{eqn:deltaHinterface}
    \end{pmatrix},\\ \nonumber
\end{align}
where $\Upsilon^{t}_{cv}$ and $\Upsilon^{t}_{v_{1}v_{2}}$ are the mixing terms between bands $c-v$ and $v_{1}-v_{2}$ in the top interface and  $-\Upsilon^{b}_{cv}$,$-\Upsilon^{b}_{v_{1}v_{2}}$ are ones mixing bands $c-v$ and $v_{1}-v_{2}$ at the bottom interface. Note that such mixing terms require an opposite sign due to the opposite sign due to the opposite direction of the interfacial effective electric fields at the two InSe/hBN interfaces. Given the very small interfacial energy shift of bands $c$ and $v$ and the very weak hybridization between bands $v_{1}$ and $v_{2}$, the dominant contribution to the conduction band SOC strength originates from the interfacial terms mixing bands of opposite parity (see Fig. \ref{fig:InterfaceFeynmann}).
Among them, the most relevant contribution originates from the term $\Upsilon_{cv}$ mixing bands $c$ and $v$, which, in the absence of an applied electric field, yields to the following contribution to the SOC strength,
\begin{align}
    &
    \Delta H^{(I)}_{11}=2\Bigg(\Big[\frac{b_{45}\lambda_{15}}{\Delta E_{g^{1}}\Delta E_{cv_{1}}}+\frac{b_{16}\lambda_{46}}{\Delta E_{g^{1}}\Delta E_{cv_{2}}}\Big](\Upsilon^{t}_{cv}\alpha_{1}\beta_{1}-\\ \nonumber
    &
    \Upsilon^{b}_{cv}\alpha_{N}\beta_{N})\Bigg)(\mathbf{\bold{s}\times k}),\\ \nonumber
\end{align}
where $\Delta E_{g^{1}}$ is the energy between the lowest conduction subband and the topmost valence band (i.e. the energy gap) and $\Delta E_{cv_{1(2)}}$ is the energy difference between the lowest conduction subband and the $v_{1(2)}$ subbands. The number of layers dependence of the interfacial SOC strength can be extracted expanding $\Delta E_{g^{1}}$, $\Delta E_{cv_{1(2)}}$, $\alpha_{1(N)}$ and $\beta_{1(N)}$ as a function of the number of layers in the quantum well approximation presented in Ref. \onlinecite{subbands2018}. In such approximate framework, the out-of-plane wavevector $k_{z}$ depends on the number of layers as $k_{z}=\frac{\pi}{a_{z}}+\frac{n\pi}{(N+2\nu)a_{z}}$ and the wavefunctions for both the conduction and the valence bands are approximated as the eigenstates of a quantum well size $L=(N+2\nu)a_{z}$ $(\Phi^{n}_{c}\approx\Phi^{n}_{v}\approx\sqrt{\frac{1}{(N+2\nu)a_{z}}}\cos\Big(\frac{n\pi}{(N+2\nu)a_{z}}\Big))$. From this quantum well model, a $\frac{1}{(N+2\nu)^{3}}$ dependence of the interfacial SOC strength is expected, as confirmed by the fit presented in Fig. \ref{fig:interfaceQW}. Given the smallness of the interfacial SOC strength compared to the layer-number dependent Dresselhaus SOC, any contribution coming from the hBN/InSe interface will be neglected for the rest of our analysis. \\

\section{L{\"o}wdin partitioning method}\label{app:Lowdin}

In order to obtain the $3^{rd}$ order corrections to the hybrid $\bold{k \cdot p}$ tight-binding Hamiltonian, the standart method of L{\"o}wdin partitioning\cite{Lowdinpart} is applied. The total multilayer Hamiltonian is written in the basis of the unperturbed subbands eigenstates obtained from diagonalizing the $\hat{H}_{0}$ part of the Hamiltonian in Eq. (\ref{eqn:H0}),
\begin{align}
    \hat{H}=\hat{H}_{0}+\hat{H}'\label{eqn:Ht}
\end{align}
where $\hat{H}'$ is the perturbative part, namely the projection of $\delta \hat{H}$ in the orthogonal subband basis formed by $\hat{H}_{0}$ ($H'_{\rho\omega}\equiv \langle \rho \vert \delta \hat{H} \vert \omega \rangle$). In the partitioning method, two diagonal blocks are defined, A and B and a unitary transformation is applied to the entire Hamiltonian matrix in order to remove the non-block-diagonal elements. The set A is defined as the elements within the lowest conduction subband $c^{1}$
\begin{align}
    \hat{H}_{A}=\begin{pmatrix}
    \langle c^{1\uparrow} |\hat{H}_{A}| c^{1\uparrow} \rangle  & \langle c^{1\uparrow} |\hat{H}_{A}| c^{1\downarrow} \rangle  \\
    \langle c^{1\downarrow} |\hat{H}_{A}| c^{1\uparrow} \rangle  & \langle c^{1\downarrow} |\hat{H}_{A}| c^{1\downarrow} \rangle \\
\end{pmatrix},
\end{align}
while the set B are the matrix elements within the valence subbands or the upper conduction subbands,
\begin{widetext}
\begin{align}
    \hat{H}_{B}=\begin{pmatrix}
    \langle v^{1\uparrow} |\hat{H}_{B}| v^{1\uparrow} \rangle  & \langle v^{1\uparrow} |\hat{H}_{B}| v^{1\downarrow} \rangle  & \langle v^{1\uparrow} |\hat{H}_{B}| v^{2\uparrow} \rangle  & \langle v^{1\uparrow} |\hat{H}_{B}| v^{2\downarrow} \rangle  \ldots \\
    \langle v^{1\downarrow} |\hat{H}_{B}| v^{1\uparrow} \rangle  & \langle v^{1\downarrow} |\hat{H}_{B}| v^{1\downarrow} \rangle & \langle v^{1\downarrow} |\hat{H}_{B}| v^{2\uparrow} \rangle & \langle v^{1\downarrow} |\hat{H}_{B}| v^{2\downarrow} \rangle  \ldots \\
    \langle v^{2\uparrow} |\hat{H}_{B}| v^{1\uparrow} \rangle & \langle v^{2\uparrow} |\hat{H}_{B}| v^{1\downarrow} \rangle &  \langle v^{2\uparrow} |\hat{H}_{B}| v^{2\uparrow} \rangle & \langle v^{2\uparrow} |\hat{H}_{B}| v^{2\downarrow} \rangle  \ldots 
    \\
    \langle v^{2\downarrow} |\hat{H}_{B}| v^{1\uparrow} \rangle & \langle v^{2\downarrow} |\hat{H}_{B}| v^{1\downarrow} \rangle & \langle v^{2\downarrow} |\hat{H}_{B}| v^{2\uparrow} \rangle & \langle v^{2\uparrow} |\hat{H}_{B}| v^{2\uparrow} \rangle \ldots \\
    \vdots & \vdots & \vdots & \vdots \\
\end{pmatrix},
\end{align}
\end{widetext}
where the numerical indices such as 1 and 2 refer to the $1^{st}$ or $2^{nd}$ subbands. The non-block-diagonal elements.
$H_{nbd}$ are the elements mixing the terms of the A and B block namely
\begin{align}
    \hat{H}_{nbd}=\begin{pmatrix}
    \langle c^{1\uparrow} |\hat{H}| v^{1\uparrow} \rangle  & \langle c^{1\uparrow} |\hat{H}| v^{1\downarrow} \rangle  & \langle c^{1\uparrow} |\hat{H}| v^{2\uparrow} \rangle & \ldots\\
    \langle c^{1\downarrow} |\hat{H}| v^{1\uparrow} \rangle  & \langle c^{1\downarrow} |\hat{H}| v^{1\downarrow} \rangle & \langle c^{1\downarrow} |\hat{H}| v^{2\uparrow} \rangle & \ldots\\
\end{pmatrix}.
\end{align}
The expression in Eq. (\ref{eqn:Ht}) is rewritten in terms of $\hat{H}'_{1}$ (the matrix containing the perturbations within block A and B), and $\hat{H}'_{2}$ (the non-zero perturbations between sets A and B)
\begin{align}
    \hat{H}=\hat{H}_{0}+\hat{H}'_{1}+\hat{H}'_{2}.
\end{align}
Transforming the Hamiltonian with a unitary transformation of the form $e^{\hat{S}}$,
\begin{align}
    \tilde{H}=e^{-\hat{S}}\hat{H}e^{\hat{S}},
\end{align}
the deeper valence band states are projected into the lowest conduction subbband. From the definition of the A block, the matrix elements $\langle c^{\uparrow}|\hat{H}|c^{\downarrow} \rangle$ and $\langle c^{\downarrow}|\hat{H}|c^{\uparrow} \rangle$ are the terms responsible for the SOC splitting. We get the following expressions for the block and non-block-diagonal matrix elements,
\begin{align}
    &
    \hat{H}_{bd}=\sum^{\infty}_{j=0}\frac{1}{(2j)!}[\hat{H}^{(0)}+\hat{H}^{(1)},\hat{S}]^{(2j)}+\nonumber \\
    &
    \quad\quad\quad \sum^{\infty}_{j=0}\frac{1}{(2j+1)!}[\hat{H}^{(2)},\hat{S}]^{(2j+1)},\nonumber \\
   &
   \hat{H}_{nbd}=\sum^{\infty}_{j=0}\frac{1}{(2j+1)!}[\hat{H}^{(0)}+\hat{H}^{(1)},\hat{S}]^{(2j+1)}+\nonumber \\
   &
   \quad\quad\quad \sum^{\infty}_{j=0}\frac{1}{(2j)!}[\hat{H}^{(2)},\hat{S}]^{(2j)}.
\end{align}
The non-block-diagonal terms are then set to 0 forcing the third order in the perturbation Hamiltonian ($\Delta H^{(3)}$) to be
\begin{widetext}
\begin{align}
    &
    \Delta H^{(3)}_{mm'}=-\frac{1}{2}\sum_{l,m''}\Big[\frac{H'_{ml}H'_{lm''}H'_{m''m'}}{(E_{m'}-E_{l})(E_{m''}-E_{l})}+\frac{H'_{mm''}H'_{m''l}H'_{lm'}}{(E_{m}-E_{l})(E_{m''}-E_{l})}\Big]\\ \nonumber
    &
    +\frac{1}{2}\sum_{l,l'}\Big[\frac{H'_{ml}H'_{ll'}H'_{l'm'}}{(E_{m}-E_{l})(E_{m}-E_{l'})}+\frac{H'_{ml}H'_{ll'}H'_{l'm'}}{(E_{m'}-E_{l})(E_{m'}-E_{l'})}\Big],
\end{align}
\end{widetext}
where ($m,m'$) are elements within A and ($l,l'$) are elements within B. Having identified the loops responsible for the SOC splitting shown in Figs. \ref{fig:gammastackedrashbac1}-\ref{fig:electrostaticU} and \ref{fig:dipole}, the mixing between the conduction and the deeper valence bands projected into the lowest conduction subband has the form,
\begin{widetext}
\begin{align}
    &
    \Delta H^{(3)}_{11}=2\sum^{j=N}_{j=1}\sum^{\eta=N}_{\eta=1}\frac{\langle c^{\downarrow} \vert \delta \hat{H}\vert v_{2,\eta}^{\uparrow}\rangle\langle v^{\uparrow}_{2,\eta}\vert \delta \hat{H} \vert v^{j,\uparrow} \rangle\langle v^{j,\uparrow} \vert \delta \hat{H} \vert c^{\uparrow} \rangle}{\Delta E_{cv_{2}}\Delta E_{cv^{j}}}+2\sum^{j=N}_{j=1}\sum^{\eta=N}_{\eta=1}\frac{\langle c^{\downarrow} \vert \delta \hat{H}\vert v_{1,\eta}^{\downarrow}\rangle\langle v^{\downarrow}_{1,\eta}\vert \delta \hat{H} \vert v^{j,\uparrow} \rangle\langle v^{j,\uparrow} \vert \delta \hat{H} \vert c^{\uparrow} \rangle}{\Delta E_{cv_{1}}\Delta E_{cv^{j}}}\\
    &
    +2\sum^{\eta=N}_{\eta=1}\frac{\langle c^{\downarrow} \vert \delta \hat{H}\vert v^{\uparrow}_{2,\eta}\rangle\langle v^{\uparrow}_{2,\eta}\vert \delta \hat{H} \vert v^{\uparrow}_{1,\eta} \rangle\langle v^{\uparrow}_{1,\eta} \vert \delta \hat{H} \vert c^{\uparrow} \rangle}{\Delta E_{cv_{1}}\Delta E_{cv_{2}}}+2\sum^{\eta=N}_{\eta=1}\frac{\langle c^{\downarrow} \vert \delta \hat{H}\vert c^{\downarrow}_{1,\eta}\rangle\langle c^{\downarrow}_{1,\eta}\vert \delta \hat{H} \vert v^{\downarrow}_{2,\eta} \rangle\langle v^{\downarrow}_{2,\eta} \vert \delta \hat{H} \vert c^{\uparrow} \rangle}{\Delta E_{cc_{1}}\Delta E_{cv_{2}}}. \nonumber
\end{align}
\end{widetext}
Knowing the origin of the 3-step loop processes described in Section \ref{section:perturbationalaysis}, the Hamiltonian that contributes to the SOC in the absence of a relevant interfacial term can be decomposed as
\begin{align}
    \Delta H^{(3)}_{11}=\Delta H'_{11}+\Delta H''_{11}+\Delta H'''_{11},
\end{align}
where the different terms correspond to the different mechanisms behind SOC in band $c$,
\begin{widetext}
\begin{align}
   &
   \Delta H'_{11}=2\Bigg[\sum^{j=N}_{j=1}\sum^{\kappa=N}_{\kappa=1}\Big( \frac{b_{54}\lambda_{15}t^{\Gamma}_{cv}}{\Delta E_{cv_{1}}\Delta E_{cv^{j}}}+\frac{b_{16}\lambda_{46}t^{\Gamma}_{cv}}{\Delta E_{cv_{2}}\Delta E_{cv^{j}}}\Bigg)\alpha_{\kappa}(\beta^{j}_{\kappa+1}-\beta^{j}_{\kappa-1})\Bigg(\sum^{\xi=N}_{\xi=1}\alpha_{l}\beta^{j}_{\xi}\Bigg)\Bigg](\mathbf{\bold{s}\times k}),\\ \nonumber
    &
    \Delta H''_{11}= 2\Bigg[\sum^{j=N}_{j=1}\sum^{\kappa=N}_{\kappa=1}\left(\frac{\mathcal{E}_{\kappa}d_{cv}\lambda_{15}b_{54}}{\Delta E_{cv^{j}}\Delta E_{cv_{1}}}+\frac{\mathcal{E}_{\kappa}d_{cv}\lambda_{46}b_{16}}{\Delta E_{cv^{j}} \Delta E_{cv_{2}}} \right)(\alpha_{\kappa}\beta^{j}_{\kappa})\Bigg(\sum^{\xi=N}_{\xi=1}\alpha_{\xi}\beta^{j}_{\xi}\Bigg)+\sum^{\eta=N}_{\eta=1}\alpha^{2}_{\eta}\bigg (\frac{\mathcal{E}_{\eta}d_{v_{1}v_{2}}b_{54}\lambda_{46}}{\Delta E_{cv_{1}} \Delta E_{cv_{2}}} +\\ \nonumber
    &
    \frac{\mathcal{E}_{\eta}d_{c_{1}c}b^{c_{1}v_{2}}_{16}\lambda_{46}}{\Delta E_{cc_{1}} \Delta E_{cv_{2}}}\bigg)  \Bigg] (\mathbf{\bold{s}\times k)},\\ \nonumber
    &
    \Delta H'''_{11}=2\Bigg[\sum^{j=N}_{j=1}\sum^{\kappa=N}_{\kappa=1}\Bigg( \frac{b_{54}\lambda_{15}\delta_{cv}}{\Delta E_{cv_{1}}\Delta E_{cv^{j}}}+\frac{b_{16}\lambda_{46}\delta_{cv}}{\Delta E_{cv_{2}}\Delta E_{cv^{j}}}\Bigg)\alpha_{\kappa}(\beta^{j}_{\kappa+1}+\beta^{j}_{\kappa-1})\Bigg(\sum^{\xi=N}_{\xi=1}\alpha_{\xi}\beta^{j}_{\xi}\Bigg)+\sum^{\eta=N}_{\eta=1}\Bigg(\frac{b^{c_{1}v_{2}}_{16}\lambda_{46}\delta_{c_{1}c}}{\Delta E_{cc_{1}}\Delta E_{cv_{1}}}\Bigg)\times\\ \nonumber 
    &
    \alpha_{\eta}(\alpha_{\eta+1}+\alpha_{\eta-1})\Bigg](\mathbf{\bold{s}\times k}). \\\nonumber
\end{align}
\end{widetext}
Finally, using Eq. (D9) the interfacial contribution to the SOC strength coming from the dominant $\Upsilon^{t/b}_{cv}$ term in Eq. (\ref{eqn:deltaHinterface}) has the form
\begin{widetext}
\begin{align}
    \Delta H^{(I)}_{11}=2\Bigg[\sum^{j=N}_{j=1}\Big(\frac{b_{45}\lambda_{15}}{\Delta E_{cv^{j}}\Delta E_{cv_{1}}}+\frac{b_{16}\lambda_{46}}{\Delta E_{cv^{j}}\Delta E_{cv_{2}}}\Big)(\Upsilon^{t}_{cv}\alpha_{1}\beta^{j}_{1}-\Upsilon^{b}_{cv}\alpha_{N}\beta^{j}_{N})\Bigg(\sum^{\xi=N}_{\xi=1}\alpha_{l}\beta^{j}_{\xi}\Bigg)\Bigg](\mathbf{\bold{s}\times k}).
\end{align}
\end{widetext}
Considering the limit where the applied electric field is zero, this term simplifies to
\begin{widetext}
\begin{align}
    \Delta H^{(I)}_{11}=-2\Big[\frac{b_{45}\lambda_{15}}{\Delta E_{g^{1}}\Delta E_{cv_{1}}}+\frac{b_{16}\lambda_{46}}{\Delta E_{g^{1}}\Delta E_{cv_{2}}}\Big](\Upsilon^{t}_{cv}\alpha_{1}\beta^{1}_{1}-\Upsilon^{b}_{cv}\alpha_{N}\beta^{1}_{N})(\mathbf{\bold{s}\times k}).\\ \nonumber
\end{align}
\end{widetext}

\bibliography{isb.bib}

\end{document}